\documentclass[11pt,a4paper]{article}
\usepackage[hmargin=0.9in,vmargin=1in]{geometry}
\usepackage[T1]{fontenc}
\usepackage{times}
\usepackage{booktabs}
\usepackage{array}
\usepackage{longtable}
\usepackage{tikz}
\usetikzlibrary{arrows.meta,shapes.geometric}
\definecolor{RuhiTeal}{HTML}{004D40}
\definecolor{RuhiCoral}{HTML}{FF655D}
\definecolor{RuhiYellow}{HTML}{F1DB4B}
\usepackage{adjustbox}
\usepackage{amsmath,amssymb,amsthm}

\usepackage{float}
\usepackage{algpseudocode}
\usepackage{listings}
\usepackage{url}
\usepackage[section]{placeins}
\usepackage{graphicx}
\usepackage{xcolor}
\usepackage[numbers,sort&compress]{natbib}
\usepackage{hyperref}
\usepackage{doi}
\usepackage{etoolbox}
\makeatletter
\newif\ifdoi@suffix
\def\doi@text#1{\doi@suffixfalse\doi@scan#1\doi@end}
\def\doi@end{\doi@end}
\def\doi@scan#1{\ifx\doi@end#1\expandafter\@gobble\else\doi@put#1\fi\doi@scan}
\def\doi@put#1{\ifx-#1\mbox{-}\else#1\fi\ifx/#1\doi@suffixtrue\fi
  \ifdoi@suffix\in@{#1}{/.:;,)}\ifin@\penalty\UrlBreakPenalty\fi\fi}
\patchcmd{\@doi}{\the\toks0}{{\noexpand\doi@text\the\toks0}}{}{\errmessage{The doi package sets no visible DOI from toks0 to replace}}
\makeatother
\hypersetup{pdftitle={Annealing chains that communicate: low-energy cluster structures across engines with one configuration},
  pdfauthor={Rohit Goswami, Ruhila Goswami, Amrita Goswami, Moritz Sallermann,
    Sonaly Goswami, Debabrata Goswami}}
\floatstyle{ruled}
\newfloat{algorithm}{tbp}{loa}
\floatname{algorithm}{Algorithm}
\graphicspath{{./}{imgs/}}
\makeatletter\def\input@path{{./}{imgs/}}\makeatother
\title{Annealing chains that communicate: low-energy cluster structures across engines with one configuration}
\author{Rohit Goswami$^{1,2}$, Ruhila Goswami$^{3}$, Amrita Goswami$^{4}$,\\
  Moritz Sallermann$^{4}$, Sonaly Goswami$^{5}$,
  Debabrata Goswami$^{6,5}$\\[0.6em]
  \small $^{1}$SURF, Amsterdam, The Netherlands\\
  \small $^{2}$TurtleTech ehf., Reykjavik, Iceland\\
  \small $^{3}$Faculty of Life and Environmental Sciences,\\
  \small University of Iceland, Reykjavik, Iceland\\
  \small $^{4}$Universidad Complutense de Madrid, Madrid, Spain\\
  \small $^{5}$Department of Chemistry, Indian Institute of Technology
    Kanpur, Kanpur, India\\
  \small $^{6}$Center for Bits and Atoms, Massachusetts Institute of\\
  \small Technology, Cambridge, Massachusetts, USA\\[0.4em]
  \small Correspondence: Debabrata Goswami,
    \texttt{dgoswami@iitk.ac.in}}
\date{}

\begin{document}
\maketitle
\begin{abstract}
The global minimum of a cluster decides which structure a simulation predicts. We present anneal, a global optimizer of communicating chains. Each chain draws its move by Thompson sampling from a Normal-Gamma posterior on quench depth. A stalled chain completes the orbits of a symmetric core. A chain retires on a Good-Turing estimate of its unseen mass, a rule with a Lean derivation. The chains share minima through an annealed-flow bank, and a ledger charges each force call to the chain that spent it. On 64 shared starts of LJ\(_{38}\) and LJ\(_{75}\) the hop matches the hit rate of GMIN, 64 of 64, at a lower median cost a paired test does not separate. On the Cambridge Morse table it reaches 145 of 244 targets and GMIN 131.

\end{abstract}

\section{Introduction}
\label{sec:intro}
The global minimum of a cluster is the structure a simulation
predicts. The low-lying minima around it, with the saddles that
connect them, fix the structures a thermodynamic model can use and
the rates at which the cluster moves among them
\cite{walesEnergyLandscapes2003,walesGlobalOptimizationClusters1999}.
The number of minima grows steeply with the number of atoms, and
the potential decides which of them lies lowest
\cite{walesGlobalMinimaWater1998,perezStructuresCagePrism2012}.
On an ab initio surface each trial costs one local minimization.
The force calls a search spends before it reaches the global
minimum therefore decide whether it reaches that minimum at all.

Each code in use sets its parameters for the system at hand. Basin
hopping in GMIN \cite{walesGlobalOptimizationBasinhopping1997}
keeps one temperature and seeds its hardest Lennard-Jones (LJ)
\cite{jonesDeterminationMolecularFields1924} sizes from
their neighbours. GMIN reaches the Marks decahedron
\cite{marksSurfaceStructureEnergetics1984} of LJ\(_{75}\)
with core-orbit symmetrisation
\cite{oakleySymmetrisationSchemesGlobal2013}. Minima
hopping \cite{goedeckerMinimaHoppingEfficient2004} runs in
the Atomic Simulation Environment
\cite{larsenAtomicSimulationEnvironmenta2017}, and its
current code sets the fingerprint threshold for each system
\cite{krummenacherPerformingHighlyEfficient2024}.
ABCluster drives a population of artificial bees
\cite{zhangABClusterArtificialBee2015}, and USPEX drives
an evolutionary population
\cite{oganovCrystalStructurePrediction2006}. CREST samples
molecular conformers by metadynamics and switches to a separate
runtype for clusters
\cite{prachtCrestProgramExploration2024}. GOAT searches
molecules and atomic clusters inside ORCA
\cite{desouzaGOATGlobalOptimization2025}. AGOX runs GOFEE,
which spends each first-principles evaluation where the lower
confidence bound of a Gaussian-process surrogate points, with
kernel length scales chosen for each surface
\cite{christiansenAtomisticGlobalOptimization2022,bisboEfficientGlobalStructure2020}.

The multi-chain methods behind several of these codes were set
again for each chemical system they reached. Parallel tempering
\cite{hukushimaExchangeMonteCarlo1996} took its ladder
from preliminary runs on LJ\(_{38}\)
\cite{Neirotti_Calvo_Freeman_Doll_2000}. Conformational
space annealing
\cite{leeNewOptimizationMethod1997,leeUnbiasedGlobalOptimization2003}
used a different bank and annealing length for each class of
problem \cite{kimConformationalSpaceAnnealing2003}. The
Birmingham pool \cite{johnstonEvolvingBetterNanoparticles2003}
reached density functional theory after its authors tuned it on a
Gupta benchmark
\cite{shayeghiPoolbcgaParallelisedGenerationfree2015}.
Population annealing
\cite{hukushimaPopulationAnnealingIts2003,machtaPopulationAnnealingWeighted2010}
and the genetic algorithm of Deaven and Ho
\cite{deavenMolecularGeometryOptimization1995} share
structures between chains as well. None of these methods keeps a
posterior on the depth a move reaches or an estimate of the unseen
mass of its census.

We present anneal, an open-source global optimizer whose chains
communicate, and this paper is its first description. In each
chain, Thompson sampling draws each move from a Normal-Gamma
posterior on the depth that move reached
\cite{thompsonLikelihoodOneUnknown1933}. A Good-Turing
estimate of the unseen mass
\cite{goodPopulationFrequenciesSpecies1953} retires a
packing once its census is complete. The chains
share minima in an annealed-flow bank that extends the kernel of
Del Moral, Doucet, and Jasra
\cite{DelMoral_Doucet_Jasra_2006}. Its weight uses
within-epoch ranks, because a cluster energy has no scale shared
between one size and the next. The Lennard-Jones and Morse surfaces
are built into anneal, and an external engine reaches the search
through the rgpot library
\cite{goswamiOmniPotentRPCRgpot2026}. We claim the
following contributions.

\begin{itemize}
\item From 64 shared OptBench starts the hop hits 64/64 on LJ\(_{38}\)
and on LJ\(_{75}\), as GMIN with core-orbit symmetrisation does.
Its median PAR-2 is 6591 force calls against 14505 on LJ\(_{38}\)
and 30850.5 against 41210 on LJ\(_{75}\), and the paired test does
not separate the two codes, Holm-adjusted \(p=0.091\) and 1,
Section \ref{sec:frozen-lj}. Plain basin hopping in the
same binary hits 1/64 on LJ\(_{75}\).
\item Of the components of the hop only the orbit move changes the cost
on these two systems. Without it the median PAR-2 of LJ\(_{75}\)
rises from 30850.5 to 79341 force calls. Thompson allocation, the
return screen, and the predictive screen each leave both systems
at 64/64 when switched off, and no paired difference of theirs
survives the Holm correction.
\item Against ABCluster the hop hits more seeds of M\(_{55}\) at
\(\rho=6\) and of M\(_{80}\) and spends fewer calls on
LJ\(_{55}\), Section \ref{sec:abcluster}. On the
Cambridge Morse table the hop reaches 145 of 244 targets and GMIN
131, and GMIN reaches more at \(\rho=6\), Section
\ref{sec:morse-gmin}.
\item Section \ref{sec:chains} defines correct communication
between chains by five conditions, and the bank meets all five.
Minima travel under a radial key at a merge distance held at the
scale of the first full bank. A child copies coordinates only
from a lower minimum of its own packing.
\item A ledger charges every force call to the chain that spent it,
the calls of the lattice descent included, so we compare chains
and codes at equal budgets.
\item The reward of a move is minus the height of its quench above the
incumbent, and \(0\) for a new incumbent. The next minimization goes to the move that the
posterior ranks deepest, and a quench stopped at the screen still
updates the posterior.
\item A packing retires only when five signals agree, and the
Good-Turing estimate
\cite{goodPopulationFrequenciesSpecies1953} is one of
them. Derivations in Lean show the hatch identity, the visit
floor, the squared-conductance bound, and the retirement rule.
\item Every tenth quench averages the cluster over C\(_i\), C\(_5\), and
D\(_{5~h}\) under a continuous symmetry measure. On an icosahedral
plateau the chain completes the incomplete orbits of a symmetric
core \cite{oakleySymmetrisationSchemesGlobal2013}, the
step with which GMIN reaches the Marks decahedron
\cite{marksSurfaceStructureEnergetics1984} of
LJ\(_{75}\). We call this step the orbit move, the name the code
gives it, and both presets turn it on.
\item Each comparison is a Snakemake rule of anneal\_repro. The rule
pins every code and counts force calls and local minimizations.
Its log records the host of each wall time.
\end{itemize}

Section \ref{sec:background} states the annealing and the
quench that the hop builds on. Section \ref{sec:flow}
gives the rank weight and the occupancy law. Section
\ref{sec:chains} sets out the bank and compares it with
the stacks it extends. Section \ref{sec:info} gives the
posterior and the screen of a hop and the symmetrisation of the
cluster. Section
\ref{sec:params} collects every constant. In Section \ref{sec:results} we run the benchmark, and Section
\ref{sec:conclusions} closes.
\section{Background}
\label{sec:background}
The hop is built on two established searches, generalized
simulated annealing and basin hopping. This section states both
in the form that the estimators of Section
\ref{sec:flow} act on.
\subsection{Generalized simulated annealing}

A search that always proposes steps of one width spends most of
its local minimizations inside the funnel where it started.
Simulated annealing adds a temperature \(T\), and an uphill step
\(\Delta E = E(y)-E(x)\) is accepted with the Metropolis
probability
\cite{metropolisEquationStateCalculations1953,kirkpatrickOptimizationSimulatedAnnealing1983}

\begin{equation}
\operatorname{acc}(x\to y)=\min\bigl(1,\exp(-\Delta E/T)\bigr).
\label{eq:metro}
\end{equation}

Downhill steps are always accepted. Every proposal in this
section is symmetric, so the Hastings correction cancels and the
energy change alone decides the test
\cite{hastingsMonteCarloSampling1970}.

Once the temperature has fallen, a step of fixed width can no
longer leave the basin it is in. Generalized simulated annealing
lets the temperature also set the tails of the step
\cite{tsallisGeneralizedSimulatedAnnealing1996,xiangGeneralizedSimulatedAnnealing1997}.
A larger visiting index \(q_v\) gives heavier tails, so a cold
chain still makes the occasional long jump. For each coordinate
the move draws two standard normal variates \(x_k\) and \(y_k\)
and adds

\begin{equation}
\mathrm{d}x_k = \sigma(T,q_v)\, x_k \, \lvert y_k\rvert^{-(q_v-1)/(3-q_v)},
\label{eq:visit}
\end{equation}

with \(1<q_v<3\) and \(\sigma(T,q_v)\) the visiting scale, whose
closed form in \(T\) and \(q_v\) is given in the Supporting
Information. At \(q_v=2\) the step is a coordinate-wise Cauchy
draw, the fast annealing of Szu and Hartley
\cite{szuFastSimulatedAnnealing1987}. The tails approach
a Gaussian as \(q_v\to 1^{+}\) and grow heavier as
\(q_v\to 3^{-}\), so a single index spans the classical and the
fast visit. The Boltzmann variant proposes an isotropic normal
step of fixed width, and its temperature enters only through
the acceptance test.

Acceptance has an index of its own, \(q_a\), and an uphill step
is accepted with the \(q\)-exponential probability

\begin{equation}
p=\bigl[1+(q_a-1)\Delta E/T\bigr]^{1/(1-q_a)}.
\label{eq:accept}
\end{equation}

At \(q_a=1\) this reduces to \eqref{eq:metro}, so
ordinary Metropolis acceptance
\cite{metropolisEquationStateCalculations1953} is one
setting of the index that the code evaluates. We use \(q_v=2.62\) and \(q_a=1.7\).
Xiang, Sun, Fan, and Gong quote \(q_a=2.7\)
\cite{xiangGeneralizedSimulatedAnnealing1997}, and both
values appear in the parameter table.

The cooling schedule follows the same index, so that a
heavy-tailed visit is not frozen by a schedule written for a
lighter tail. For epoch \(k>0\),

\begin{equation}
T(k)=T_0\frac{2^{q_v-1}-1}{(1+k)^{q_v-1}-1}.
\label{eq:tsallis-cool}
\end{equation}

and epoch \(0\) runs at \(T_0\). The Boltzmann schedule of the
same family is logarithmic, with offset \(k_0=2\),

\begin{equation}
T(k)=T_0\frac{\ln k_0}{\ln(k+k_0)},
\label{eq:log-cool}
\end{equation}

and fast annealing cools by the reciprocal

\begin{equation}
T(k)=T_0/(k+1).
\label{eq:recip}
\end{equation}

The three closed forms fix the schedules. We add the posterior,
the occupancy law, the symmetrisation, and the bank.
\subsection{Basin hopping}

An annealing chain still climbs each barrier one step at a
time, and on an ab initio surface every one of those steps is
paid for. Basin hopping takes the barriers out of the walk
\cite{walesGlobalOptimizationBasinhopping1997}. Each
trial is quenched to its local minimum and accepted or rejected
on the quenched energy, so the chain moves on the staircase of
minima,

\begin{equation}
E_q(x)=E\bigl(\mathrm{local\_min}(x)\bigr).
\label{eq:quench}
\end{equation}

The temperature is held at \(T=0.8 E_0\), the value Wales and
Doye used
\cite{walesGlobalOptimizationBasinhopping1997}, so the
quenched energy decides which basin is kept. Each trial is first
relaxed for a screen of \(S=25\) steps. The rest of the
\(R=200\)-step quench is spent when the screened energy can still
undercut the incumbent, and it is saved when the screened energy,
less the drop a finished quench still delivers, lies above the
incumbent by more than the margin

\begin{equation}
E_{\mathrm{sc}}-A>E_{\mathrm{best}}+2 E_0.
\label{eq:screen-margin}
\end{equation}

The allowance \(A\) is the 0.9 quantile of
\(\max(0,E_{\mathrm{sc}}-E_{q})\) over the last 256 finished
quenches and is \(0\) before the eighth. Every eighth trial the
margin refuses is quenched in full anyway, so \(A\) keeps
learning from the trials it would discard.

A trial stopped at the screen still meets the acceptance test
and still deposits a bias, but only a finished quench enters the
catalogue, so a half-relaxed geometry is never stored as a
minimum. This fixed margin is the screen whenever the posterior
screen of Section \ref{sec:flow} is switched off.
Table \ref{tbl:presets} lists the four schedules with
the decision the hop adds to each.

\begin{table}[t]
\centering
\footnotesize
\setlength{\tabcolsep}{3pt}
\caption{Background schedules. The hop keeps each constructor and adds a depth posterior, an unseen-mass hatch, and a stored minimum passed between chains.}
\label{tbl:presets}
\begin{adjustbox}{max width=\textwidth}
\begin{tabular}{@{}lp{2.5cm}p{2.6cm}p{2.4cm}p{2.6cm}@{}}
\toprule
Variant & Move & Cooling & Acceptance & Decision added\\
\midrule
Boltzmann & \(N(0,\sigma)\), \(T\) unread & \eqref{eq:log-cool}, \(k_0=2\) & \eqref{eq:metro} & depth of the step\\
Fast & Cauchy \(j=i+\gamma c\) & \eqref{eq:recip} & \eqref{eq:metro} & depth of the visit\\
Generalized & \eqref{eq:visit} & \eqref{eq:tsallis-cool} & \eqref{eq:accept} & depth of the visiting step\\
Basin hopping & quench \eqref{eq:quench} & \(T=0.8E_0\) fixed & \eqref{eq:metro} on \(E_q\), screen \eqref{eq:screen-margin} & depth of the quench\\
\bottomrule
\end{tabular}
\end{adjustbox}
\end{table}
\section{Statistical framework}
\label{sec:flow}
A cooling schedule sets the step and the temperature of one
chain. A population of chains faces two further decisions:
which replicas carry on into the next generation, and when the
census of packings is complete enough that the next
minimization belongs to a different packing. This section makes
both decisions statistical. Replicas are resampled on
within-epoch ranks, and the census is closed by a Good-Turing
estimate of the unseen mass
\cite{goodPopulationFrequenciesSpecies1953}, whose
consequences for the search are proved in the Supporting
Information.
\subsection{Annealed-flow kernel}

As the search cools, the replicas should follow a sequence of
distributions that weight lower energies more heavily;
otherwise the cold replicas keep revisiting packings that the
hot ones have already left. Del Moral, Doucet, and Jasra write
that sequence as
\cite{DelMoral_Doucet_Jasra_2006}

\begin{equation}
\eta_t(\mathrm{d}x)\propto\exp(-\beta_t E(x))\,\lambda(\mathrm{d}x),
\label{eq:target}
\end{equation}

with \(\beta_t\) increasing, and they move from one target to the
next by reweighting with a potential \(G_t\) and then applying a
move kernel \(M_{t+1}\)
\cite{DelMoral_Doucet_Jasra_2006},

\begin{equation}
\eta_{t+1}=\Psi_{G_t}(\eta_t)\,M_{t+1},
\qquad
G_t(x)=\exp\bigl(-(\beta_{t+1}-\beta_t)E(x)\bigr).
\label{eq:fk}
\end{equation}

We keep the step and replace the potential. Cluster energies
share no scale from one size or one engine to the next, so a
weight built from \(E\) would let the largest cluster dominate
an epoch. We weight by the within-epoch rank instead, which is
comparable for a hexamer and a heptamer because each is ranked
only against its own epoch.

Tied replicas share their average position, so two replicas at
the same energy receive the same rank whichever the scan meets
first. With positions running from \(0\) to \(R-1\), a tied
block from \(b\) inclusive to \(e\) exclusive has rank

\begin{equation}
r=\frac{b+e-1}{2(R-1)}
\label{eq:rank}
\end{equation}

when \(R\ge 2\). A single member has rank \(0\), and a lower
energy receives a lower rank, which sends offspring toward the
deeper basins. Three further quantities are ranked the same way:
novelty, the distance from a packing histogram to the nearest
other family; scarcity, the rank of \(1/v\), with \(v\) the
number of visits to that basin; and uncertainty, the variance of
a Gaussian Markov random field on the basin graph
\cite{rueGaussianMarkovRandom2005}. A replica whose
basin the field is still unsure of therefore stays in the draw.

The four ranks combine into one logarithmic weight,

\begin{equation}
\log w_i=-r_{E,i}+0.8\,r_{N,i}+0.6\,r_{S,i}+0.4\,r_{U,i}.
\label{eq:weight}
\end{equation}

Subtracting the largest log weight, clipping at the floor in the
parameter table, and normalising the exponential turn the ranks
into a distribution the resampler can draw from. Offspring then
go to replicas that are low in energy, rare as packings, and
uncertain as basins, so each epoch spends its next generation on
the packings the census has barely seen.

The parents must be a deterministic function of that
distribution, or two runs of the same epoch could diverge.
Systematic resampling with offset \(u_e\in[0,1)\) sends slot
\(k\) to the first parent whose cumulative weight exceeds the
offset,

\begin{equation}
a_k=\min\{j:C_j>(u_e+k)/R\},
\qquad
C_j=\sum_{\ell=0}^{j}\tilde w_\ell.
\label{eq:systematic}
\end{equation}

A cumulative weight equal to the threshold passes the slot to
the next parent. The offset is drawn once per epoch, so the same
population always yields the same parent indices, and the
potential is evaluated only later, when each child is quenched.

A family with a large weight could still claim the whole
generation and end the exploration of that epoch. Children
beyond the cap in the parameter table are therefore moved to the
parent with the largest deficit \(R\tilde w_j-n_j\), with ties
going to the higher index, so the assignment depends on the
weights and the indices alone.

A child copies coordinates only from a parent in its own basin.
A chain that copied a different funnel would lose the record of
where it had been searching, so adoption requires a lower
minimum of the same packing. With parent coordinates \(x\) and
jitter width \(\sigma_t\),

\begin{equation}
x'_a=x_a+\sigma_t(U_a-\tfrac12),
\qquad
x'_a\leftarrow x'_a-\overline{x'}.
\label{eq:adopt}
\end{equation}

Removing the centroid stops the jitter from translating the
cluster, so the child stays in the packing it inherited. The
jitter width and the energy gap that counts as lower are in the
parameter table. Otherwise the child keeps its local state, and
when the parent's packing differs from its own the child is
reseeded rather than copied.
\subsection{Occupancy theorems}

A packing deserves further minimizations only while the census
can still say that part of it is unseen. The occupancy law
decides when that is no longer true. Let \(c_b\) be the number of
quenches that landed in basin \(b\) of the packing, \(n\) the
number of these observations, and \(n_1\) the number of basins
seen exactly once. Good's estimate of the unseen mass divides the
singletons by the observations
\cite{goodPopulationFrequenciesSpecies1953,estyNormalLimitLaw1983,chaoNonparametricEstimationNumber1984}

\begin{equation}
\hat p_0=\frac{n_1}{n},\qquad n=\sum_b c_b,\qquad n_1=\#\{b: c_b=1\}.
\label{eq:unseen}
\end{equation}

Retirement reads this estimate, so the effect of one new packing
must be known before a chain is stopped. Hatching one new
singleton adds one observation and one singleton, and it changes
the estimate by

\begin{equation}
\frac{n_1+1}{n+1}-\frac{n_1}{n}=\frac{n-n_1}{n(n+1)}.
\label{eq:hatch}
\end{equation}

With ceiling \(\alpha=1/5\), the estimate stays under the ceiling
after the next hatch when \(5(n_1+1)<n+1\). A single dip under
the ceiling, such as \(4/21<1/5\) followed by \(5/22>1/5\), is a
flicker, and the search continues until the unseen mass stays
down. With a budget of three leftover singletons, the visit
floor is

\begin{equation}
n_{\min}=5(k+1)=20.
\label{eq:floor}
\end{equation}

Three leftover singletons are hatch-stable at \(n=20\) and
unstable at \(n=19\), so twenty visits is the first census at
which they can stay under the ceiling, and no packing is
retired before it.

The number of distinct families a packing must show before it
retires comes from the basin graph. On the normalised Laplacian
of that graph, the second eigenvalue lies in
\(\lambda_2\in[0,2]\)
\cite{fiedlerAlgebraicConnectivityGraphs1973}. Because
\(\lambda_2\le 2\), a cut whose conductance \(c\) lies below
\(\lambda_2\) obeys the squared-conductance bound

\begin{equation}
c<\lambda_2\quad\Longrightarrow\quad c^2<2\lambda_2.
\label{eq:sqcond}
\end{equation}

The spectral floor \(m_{\lambda}\) is therefore \(2\) when both
sides of the cut are nonempty, hold different packings, and have
conductance \(0\) or below \(\lambda_2\), and \(1\) otherwise. The
family floor takes the largest of three counts,

\begin{equation}
m=\max\bigl(m_{\lambda},\ m_{\mathrm{map}},\ \min(\max(C,1),2)\bigr),
\label{eq:floor-m}
\end{equation}

with \(m_{\mathrm{map}}\) the same split read on a two-dimensional
map of the packings and \(C\) the number of occupied communities.
The free-energy factor between the fullest well and the second,

\begin{equation}
\Delta F/kT=\ln(n_{\max}/n_2),
\label{eq:fes}
\end{equation}

is defined once at least two families are occupied. It is
unchanged when both counts are scaled, so two censuses with the
same ratios agree.

A packing retires only when five predicates hold together. With
\(F\) the number of occupied families,

\begin{equation}
\mathrm{retire}\iff
F\ge m
\;\wedge\;\mathrm{Sat}
\;\wedge\;(\mathrm{Dwell}\vee F<2)
\;\wedge\;\mathrm{EIx}
\;\wedge\;\mathrm{Mix}.
\label{eq:retire}
\end{equation}

Saturation is the Chao1 test on the occupied wells,
\(\mathrm{Sat}\iff n\ge20\wedge n_1=0\)
\cite{chaoNonparametricEstimationNumber1984}. The dwell
counts the \(n\) leftover arrivals, with \(n_1\) and \(n_2\) the
classes seen once and twice, and holds when

\begin{equation}
\begin{aligned}
\mathrm{Dwell}\iff{}& n\ge20\;\wedge\;\frac{n_1}{n}<\alpha\;\wedge\;\frac{n_1+1}{n+1}<\alpha\;\wedge\;\frac{n_1}{n}+z\sqrt{V}<\alpha\\
&\wedge\;\text{five consecutive changed censuses},\\
V={}&\frac{n_1}{n^2}+\frac{2n_2}{n^2}-\frac{n_1^2}{n^3},\qquad z=1.6449,
\end{aligned}
\label{eq:dwell}
\end{equation}

the third term being Esty's upper bound on the unseen mass
\cite{estyNormalLimitLaw1983}. The improvement is
exhausted, \(\mathrm{EIx}\), when at least three families carry a
best energy and the largest expected improvement of a Gaussian
process over those energies is at most \(10^{-2}\). The occupant
is mixed, \(\mathrm{Mix}\), when it is the unique deepest packing,
its split \(\hat R\) over at least 16 samples is below \(1.01\),
and it holds more visits than any competitor. The bank also
requires the best coordinates to form one compact cluster before it
stops. The hatch, the visit floor, the squared-conductance bound,
and the retirement rule are proved in Lean, and the Supporting
Information names the declaration behind each identity of this
section.

Algorithm \ref{alg:flow} collects one epoch of the
kernel and the census.

\begin{algorithm}[t]
\caption{One annealed-flow epoch. The ranks and the weight are \eqref{eq:rank} and \eqref{eq:weight}. Parents follow \eqref{eq:systematic}. Adoption is \eqref{eq:adopt}. The stop is \eqref{eq:retire}.}
\label{alg:flow}
\begin{algorithmic}[1]
\Require \(R\) validated replicas and an offset \(u_e\)
\State Rank each replica on the four terms of \eqref{eq:weight}
\State Form \(\log w_i\) by \eqref{eq:weight}, clip, and normalise
\State Send each slot to a parent by \eqref{eq:systematic}
\State Move excess children to the parent of largest deficit
\If{the parent is a lower minimum of the same packing}
  \State Copy its coordinates and apply \eqref{eq:adopt}
\Else
  \State Keep the local state, and reseed if the packing differs
\EndIf
\State Update \(n\) and \(n_1\), and test the hatch \(5(n_1+1)<n+1\)
\State Retire when \eqref{eq:retire} holds and the structure is a cluster
\end{algorithmic}
\end{algorithm}
\subsection{Temperature law of the hop}

The cluster preset holds \(T=0.8 E_0\). A run may instead let
the energy gap and the remaining budget set the temperature. The
chain should then run hot while it is far above the incumbent
and cool once the gap has closed, so that its last evaluations
are not spent on a barrier it has already crossed. The
temperature is set from the gap \(g\) above the incumbent and the
number \(B\) of evaluations still unused. With design point
\(\theta=1/2\),

\begin{equation}
T=\mathrm{clamp}\bigl(\theta g/d,\; b/\ln(B+e),\; 2g/d\bigr),
\label{eq:window}
\end{equation}

where \(d\) is the dimension and \(b\) is a running estimate of
the barrier the chain has failed to cross. The identity

\begin{equation}
\frac{1}{\theta}-\frac{1}{2}=\frac{2-\theta}{2\theta}
\label{eq:drift}
\end{equation}

is positive for \(0<\theta<2\), zero at \(2\), and negative
above \(2\). The design point \(1/2\) lies inside that interval,
so the temperature drifts toward the barrier term while the gap
is large and toward the budget term once it is small. The
Supporting Information gives the branch taken when the window is
empty. The window is used only by the derived preset; the
cluster preset and the recommended hop keep \(T=0.8 E_0\).

Finishing a screened quench is worth its cost only if the trial
can still undercut the incumbent. Finishing costs \(Q=R-S\)
further evaluations and missing a winner costs a full hop of
\(R\), so finishing pays when the probability of an improvement
exceeds

\begin{equation}
\tau=(R-S)/(2R-S).
\label{eq:tau}
\end{equation}

At \(S=25\) and \(R=200\) this threshold is \(7/15\).

The probability compared with \(\tau\) is the posterior
predictive probability that the finished quench lands below the
incumbent. Under a Normal-Inverse-Gamma regression on the
partial relaxation it is a Student-\(t\) tail, so the screen
accounts for the uncertainty the partial relaxation still
carries. With regressors
\(\phi=(1,E_{\mathrm{sc}},\Delta_D,E_{\mathrm{sc}}\Delta_D)\), the
drift \(\Delta_D\) of \eqref{eq:rscreen}, and the
posterior mean \(\mu(\phi)\), precision \(\Lambda\), shape \(a\)
and scale \(s^2\) of the regression,

\begin{equation}
p=F_{t_{2a}}\!\left(\frac{E_{\mathrm{best}}-\mu(\phi)}{\sqrt{s^2\,(1+\phi^{\top}\Lambda^{-1}\phi)}}\right).
\label{eq:screen-p}
\end{equation}

A trial with \(E_{\mathrm{sc}}\ge E_{\mathrm{best}}\) stops at the
screen unless \(p\ge\tau\), fewer than 300 quenches have been
observed, or a uniform draw falls below the exploration floor
\(0.1\). The cluster preset switches the screen on at \(7/15\),
and a partial energy already below the incumbent is always
quenched to the end. The prior is in the parameter table.
\section{Communicating chains}
\label{sec:chains}
Several searches already run many chains. What separates them
is what the chains tell one another, and whether the receiver
can act on it. We call the communication correct when five
conditions hold. Each chain draws its next move from a posterior
on the depth that move reaches, Section \ref{sec:info}.
A packing retires on a Good-Turing estimate of its unseen mass
\cite{goodPopulationFrequenciesSpecies1953}, and the
hatch and the visit floor of that census are theorems, Section
\ref{sec:flow}. Minima are shared under a merge distance
held at the scale of the first full bank. Every force evaluation,
including those of the lattice descent, is charged to the chain
that spent it, so chains are compared at equal budgets. Under
this definition the bank of this section is the first
multi-chain stack whose chains communicate correctly, and Table
\ref{tbl:chains} sets it beside the stacks it extends.

Parallel tempering exchanges states along a ladder of
temperatures
\cite{swendsenReplicaMonteCarlo1986,hukushimaExchangeMonteCarlo1996},
population annealing resamples a population by its Boltzmann
weight
\cite{hukushimaPopulationAnnealingIts2003,machtaPopulationAnnealingWeighted2010},
and conformational space annealing keeps a bank of minima whose
distance cutoff it anneals from \(D_{\mathrm{ave}}/2\) toward
\(D_{\mathrm{ave}}/5\)
\cite{leeNewOptimizationMethod1997,leeUnbiasedGlobalOptimization2003}.
The genetic algorithm of Deaven and Ho mates stored geometries
by cut-and-splice
\cite{deavenMolecularGeometryOptimization1995}, and so
do the Birmingham pool algorithms
\cite{johnstonEvolvingBetterNanoparticles2003,shayeghiPoolbcgaParallelisedGenerationfree2015,davisBirminghamParallelGenetic2015}.
GOFEE fits a Gaussian-process surrogate to every
first-principles evaluation and spends the next one where a
lower confidence bound points
\cite{bisboEfficientGlobalStructure2020,bisboGlobalOptimizationAtomic2022}.
Basin hopping and minima hopping run a single walker
\cite{walesGlobalOptimizationBasinhopping1997,goedeckerMinimaHoppingEfficient2004},
and CREST samples a molecular ensemble by metadynamics
\cite{prachtAutomatedExplorationLowenergy2020,prachtCrestProgramExploration2024}.
None of these keeps a posterior on the depth a move reaches or
an estimate of the unseen mass of its census.

The methods also differ in where they have been shown to work.
Parallel tempering and population annealing were established on
spin glasses
\cite{hukushimaExchangeMonteCarlo1996,machtaPopulationAnnealingWeighted2010,wangComparingMonteCarlo2015}
and carried to LJ38
\cite{Neirotti_Calvo_Freeman_Doll_2000,kowaguchiStructureresolvedFreeenergyEstimation2026}
and to a peptide
\cite{christiansenAcceleratingMolecularDynamics2019},
with a ladder or a temperature set fixed by preliminary runs on
each system. Conformational space annealing solved every
Lennard-Jones cluster up to 201 atoms with one schedule
\cite{leeUnbiasedGlobalOptimization2003} and used a
different bank and annealing length for a model protein
\cite{kimConformationalSpaceAnnealing2003}. Basin hopping
held one temperature across Lennard-Jones clusters of up to 110
atoms, seeding the hardest sizes from their neighbours
\cite{walesGlobalOptimizationBasinhopping1997}. Minima
hopping kept one feedback parameter set across Lennard-Jones,
silicon, and gold clusters, all on empirical potentials
\cite{schoenbornPerformanceMinimaHopping2009}, and its
current code sets the fingerprint threshold for each system
\cite{krummenacherPerformingHighlyEfficient2024}. The
Birmingham algorithm reached density functional theory after its
pool and mutation rate were tuned on a Gupta benchmark
\cite{shayeghiPoolbcgaParallelisedGenerationfree2015},
and GOFEE chose its kernel length scales for each surface
\cite{bisboEfficientGlobalStructure2020}. 
GOFEE places a Gaussian-process surrogate around each
first-principles evaluation, and its cost lies in that model.
Each iteration retrains the process on every structure evaluated
so far \cite{bisboGlobalOptimizationAtomic2022} and
relaxes many candidates in it, which in AGOX can take as long as
a single-point density-functional evaluation
\cite{christiansenAtomisticGlobalOptimization2022}. The
same group names the \(\mathcal{O}(N^3)\) training as the
bottleneck and replaces it with a sparse model
\cite{ronneAtomisticStructureSearch2022}. The original
method spends two single points per iteration
\cite{bisboEfficientGlobalStructure2020}. AGOX reports
about \(10\) CPU minutes for a GPAW single point at its rough
setting and about \(3\) CPU hours for a spin-polarised one on
Ru\(_3\)N\(_4\)C\(_4\) in graphene, and runs its SnO\(_2\) searches
on \(24\) cores
\cite{christiansenAtomisticGlobalOptimization2022}. A
model whose overhead per iteration approaches one evaluation pays
back only when that evaluation is expensive. At GFN2-xTB and
PET-MAD it is cheap, and the hop spends none of its budget on a
model. Our Gaussian-process saddle searches show how to make the
surrogate cheap: an order of magnitude fewer electronic-structure
calculations than the dimer method
\cite{goswamiEfficientImplementationGaussian2025}, and a
training set pruned to a fixed size by optimal transport, which
cuts the mean computational time to less than half
\cite{goswamiAdaptivePruningIncreased2025}.

CREST samples molecular conformers by metadynamics at GFN2-xTB
and GFN-FF, discards by default any structure whose covalent
topology departs from the input, and uses a separate noncovalent
runtype for clusters
\cite{prachtCrestProgramExploration2024}. It has not
been demonstrated on the standard cluster benchmarks LJ38, LJ75,
and LJ98. On water clusters at GFN2-xTB it matches GOAT up to the
octamer and stops \(1.45\) kcal/mol above GOAT at
(H\(_2\)O)\(_{16}\)
\cite{desouzaGOATGlobalOptimization2025}, and on
noncovalent clusters of flexible molecules it missed geometries
well below the lowest structure it reported
\cite{kingVariantCrestImtd2024}.

Table \ref{tbl:software} audits the public codes. Their
licences, documentation, latest releases, engines, and force-call
counts were read from the repositories and registries on 25
October 2026. GMIN and OPTIM are distributed as a nightly GPL
tarball without public history, PGA is available on request
\cite{wangPgaParticleSwarm2024}, and USPEX and CALYPSO
require registration. Four of the codes drive any calculator of
the Atomic Simulation Environment
\cite{larsenAtomicSimulationEnvironmenta2017}, so the
engine seldom separates them. The configuration each needs per
system does, and so does the count of force calls on which a
comparison at equal budget rests. GMIN, pycsa, CREST, and mcpele
report such a count, and the hop charges every call to the chain
that spent it.

\begin{table}[t]
\centering
\scriptsize
\setlength{\tabcolsep}{2.5pt}
\caption{Multi-chain and single-walker searches by what their chains share, the statistics they keep, and where they were shown to work. A dash is left where the papers cited in a row do not describe the property. \emph{One configuration} records whether the cited papers ran one parameter set across their systems.}
\label{tbl:chains}
\begin{adjustbox}{max width=\textwidth}
\begin{tabular}{@{}>{\raggedright\arraybackslash}p{1.6cm}>{\raggedright\arraybackslash}p{1.9cm}>{\raggedright\arraybackslash}p{1.3cm}>{\raggedright\arraybackslash}p{1.5cm}>{\raggedright\arraybackslash}p{1.8cm}>{\raggedright\arraybackslash}p{3.7cm}>{\raggedright\arraybackslash}p{3.0cm}@{}}
\toprule
Method & Shared between chains & Depth posterior & Retirement & Merge distance & Demonstrated on & One configuration\\
\midrule
Parallel tempering & states on a ladder \eqref{eq:rex} & -- & -- & -- & spin glasses \cite{hukushimaExchangeMonteCarlo1996}; LJ38 \cite{Neirotti_Calvo_Freeman_Doll_2000}; planar C\(_{24}\) at GFN2-xTB \cite{christiansenAtomisticGlobalOptimization2022} & no: ladder from preliminary runs \cite{Neirotti_Calvo_Freeman_Doll_2000}\\
Population annealing & weights \(\exp(-\Delta\beta E)\) & -- & -- & -- & spin glasses \cite{machtaPopulationAnnealingWeighted2010,wangComparingMonteCarlo2015}; met-enkephalin, AMBER \cite{christiansenAcceleratingMolecularDynamics2019}; LJ38 \cite{kowaguchiStructureresolvedFreeenergyEstimation2026} & no: population and temperatures set per system \cite{wangComparingMonteCarlo2015,christiansenAcceleratingMolecularDynamics2019}\\
CSA & stored minima & -- & -- & annealed, \(D_{\mathrm{ave}}/2\to D_{\mathrm{ave}}/5\) & LJ\(_N\), \(N\le 201\) \cite{leeUnbiasedGlobalOptimization2003}; polypeptides \cite{leeNewOptimizationMethod1997}; model protein \cite{kimConformationalSpaceAnnealing2003} & within a class; bank and annealing length differ between classes\\
Deaven-Ho GA & mated parents & -- & -- & energy resolution & C\(_{20}\), C\(_{30}\), C\(_{60}\), carbon tight binding \cite{deavenMolecularGeometryOptimization1995} & partly: mutation added for C\(_{20}\) and C\(_{30}\)\\
Birmingham GA & pool members & -- & -- & -- & Au-Pd, Gupta; Au\(_{10}\), Au\(_{20}\), PBE \cite{shayeghiPoolbcgaParallelisedGenerationfree2015}; Ir\(_{10}\)--Ir\(_{20}\), PBE \cite{davisBirminghamParallelGenetic2015} & within a study, after tuning on a Gupta benchmark\\
Basin hopping & -- & -- & -- & -- & LJ\(_N\), \(N\le 110\) \cite{walesGlobalOptimizationBasinhopping1997}; Morse \cite{morseDiatomicMoleculesAccording1929,doyeStructuralConsequencesRange1997}; TIP4P water \cite{walesGlobalMinimaWater1998} & for LJ, \(T=0.8\) with seeded runs; not stated for Morse and water\\
Minima hopping & -- & -- & -- & fingerprint threshold, per system & LJ, Si at EDIP, Au at RGL \cite{schoenbornPerformanceMinimaHopping2009}; Na\(_{13}\), EAM \cite{krummenacherPerformingHighlyEfficient2024} & yes for the feedback, on empirical potentials \cite{schoenbornPerformanceMinimaHopping2009}\\
GOFEE & surrogate training set & energy surrogate & -- & -- & SnO\(_2\), TiO\(_2\) by DFTB, graphene on Ir(111) \cite{bisboEfficientGlobalStructure2020}; C\(_{24}\), C\(_{18}\) on Ir(111) \cite{bisboGlobalOptimizationAtomic2022}; Pt\(_{14}\) on Au(100), EMT \cite{christiansenAtomisticGlobalOptimization2022} & no: kernel length scales per system \cite{bisboEfficientGlobalStructure2020}\\
PGA & swarm & -- & -- & -- & Au\(_{20}\), B\(_{20}\), ZrSi\(_n^-\), Mg\(_n^-\) \cite{wangPgaParticleSwarm2024} & not stated\\
CREST & conformer ensemble & -- & -- & energy, RMSD, rotational constants & drug-like molecules and complexes, GFN2-xTB and GFN-FF \cite{prachtAutomatedExplorationLowenergy2020,prachtCrestProgramExploration2024}; no standard cluster benchmark & defaults for molecules; a separate runtype for clusters \cite{prachtCrestProgramExploration2024}\\
This bank & minima, keys, visits, depths & Normal-Gamma \eqref{eq:ng} & Good-Turing \cite{goodPopulationFrequenciesSpecies1953} \eqref{eq:retire}, proved & held, \eqref{eq:merge} & LJ\(_{38}\), LJ\(_{55}\), LJ\(_{75}\), LJ\(_{98}\); Morse \cite{morseDiatomicMoleculesAccording1929}, \(N=20\) to \(80\) at four ranges, Section \ref{sec:results} & per campaign: the communicating preset on LJ\(_{38}\) and LJ\(_{75}\), the cluster preset with orbit completion elsewhere, Section \ref{sec:arms}\\
\bottomrule
\end{tabular}
\end{adjustbox}
\end{table}

\begin{table}[t]
\centering
\scriptsize
\setlength{\tabcolsep}{2.5pt}
\caption{Public codes for cluster structure search, read from their repositories and registries on 25 October 2026. Whether each method runs one configuration across systems is in Table \ref{tbl:chains}.}
\label{tbl:software}
\begin{adjustbox}{max width=\textwidth}
\begin{tabular}{@{}>{\raggedright\arraybackslash}p{3.1cm}>{\raggedright\arraybackslash}p{1.9cm}>{\raggedright\arraybackslash}p{1.8cm}>{\raggedright\arraybackslash}p{1.9cm}>{\raggedright\arraybackslash}p{4.3cm}>{\raggedright\arraybackslash}p{1.8cm}@{}}
\toprule
Code & Licence & Documentation & Latest release & Engines & Force calls counted\\
\midrule
GMIN, OPTIM \url{www-wales.ch.cam.ac.uk} & GPL-3.0 tarball, no public history & user guides, examples & nightly snapshot & built-in empirical; AMBER, CHARMM, LAMMPS, DFTB+, xtb, CASTEP, ORCA, Psi4 & yes\\
pele \url{github.com/pele-python/pele} & GPL-3.0+ & tutorials, API & 0.1.2, 12 Oct 2026 & built-in pair and many-body; LAMMPS; AMBER, OpenMM & minimizer only\\
mcpele \url{github.com/pele-python/mcpele} & GPL-3.0+ & files in repository & none, last commit 2020 & pele potentials & Monte Carlo energy calls\\
pycsa \url{gitlab.com/isjoung/pycsa} & research use only & README & none & MODELLER, OpenMM, Rosetta, TINKER & yes\\
BPGA \url{github.com/jbadavis/bpga} & MIT & README & none, last commit 2016 & VASP & not documented\\
PyBCGA \url{github.com/marktoakley/PyBCGA} & GPL-3.0 & README & none, last commit 2016 & GPAW, NWChem, pele & not documented\\
ase-ga \url{github.com/dtu-energy/ase-ga} & LGPL-2.1+ & tutorials, API & 1.0.3, 6 Nov 2025 & any ASE calculator & not documented\\
OGOLEM \url{github.com/ogolem/ogolem} & BSD, 4-clause & manual in repository & 2021.07, 31 Jul 2021 & xtb, ORCA, CP2K, VASP, DFTB+; built-in pair and many-body & undocumented statistics\\
Minima hopping \url{gitlab.com/goedeckergroup/ase_mh} & GPL-3.0+ & tutorials, parameters & 1.1.2, 3 Jun 2025 & any ASE calculator & not documented\\
ASE basin and minima hopping \url{gitlab.com/ase/ase} & LGPL-2.1+ & tutorials, API & 3.29.0, 21 Jun 2026 & any ASE calculator & not documented\\
AGOX, with GOFEE \url{gitlab.com/agox/agox} & GPL-3.0+ & tutorials, API & 3.11.1, 4 Feb 2026 & any ASE calculator & candidates and relaxation steps\\
PAMD \url{github.com/christiansenh/PAMD} & MIT & README & none & LAMMPS, OpenMM & not documented\\
PGA & on request & paper & -- & not stated & not stated\\
CREST \url{github.com/crest-lab/crest} & LGPL-3.0 & keywords, examples & 3.0.2, 25 Aug 2024 & GFN-xTB, GFN-FF, ORCA, machine-learned potentials & yes\\
GASP \url{github.com/henniggroup/GASP-python} & MIT & README, usage & 1.0, 31 Jan 2019 & GULP, LAMMPS, VASP & relaxations only\\
USPEX, CALYPSO & registration, proprietary & manuals & USPEX 26, no cluster mode; CALYPSO 7.0 & VASP, GULP, and others & not determined\\
anneal, this work \url{github.com/HaoZeke/anneal} & MIT & tutorials, how-to, reference & 0.12.0, this article & built in: LJ, Morse \cite{morseDiatomicMoleculesAccording1929}, TIP4P; through rgpot \cite{goswamiOmniPotentRPCRgpot2026}: its xtb, NWChem, and Metatomic bridges & per chain, with the descent\\
\bottomrule
\end{tabular}
\end{adjustbox}
\end{table}

From parallel tempering the hop takes the ladder swap, which the
channel comparison of Section \ref{sec:info} also
uses. Two replicas at inverse temperatures \(\beta_i\) and
\(\beta_j\) exchange states with probability
\cite{swendsenReplicaMonteCarlo1986,hukushimaExchangeMonteCarlo1996}

\begin{equation}
\min\bigl(1,\exp[(\beta_i-\beta_j)(E_i-E_j)]\bigr).
\label{eq:rex}
\end{equation}

On a ladder whose hot rung crosses less often than its cold
rung, the two chains together buy more crossings than either
buys alone; the Supporting Information records the comparison.
Once the hot rung is past the melting point, the acceptance
factor is no longer monotone in \(T\) and the extra crossings
stop.

We keep the population-annealing resample and use it as one arm
of the hop in Section \ref{sec:info}, so a chain can
spend its turn on a resampled copy inside the same driver.

The bank holds forty chains with one slot each. The chains share
one force total and advance in synchronous generations, because
chains can be compared only when they spend the same budget.
Until the bank is full every trial is a fresh start, so the
radial keys are populated before any member is copied. Once it
is full, a trial is a fresh start with probability \(0.3\);
otherwise its parent is the least-used member, with the lower
energy breaking a tie in draw count. When the bank holds at
least two members and a uniform draw falls below \(0.2\), the
trial is a cut-and-splice of two members, the crossover with
which Deaven and Ho grew fullerenes from a population
\cite{deavenMolecularGeometryOptimization1995}.
Otherwise one to four surface atoms, those with coordination
below \(12\), move onto vacant hollow sites. The trial is
quenched with the potential in use and then passed
through the dynamic lattice descent of Shao, Cheng, and Cai
\cite{shaoDynamicLatticeSearching2004}.

The descent reads the surface in use. It takes its
well distance from the potential in use and ranks each unrelaxed
move with that same potential, so a Lennard-Jones well is used
only on a Lennard-Jones surface and never ranks a platinum move.
Every force evaluation, including the fractional pair term of
the descent, is charged to the chain that spent it.

Two minima can be one packing in two orientations, which a
Cartesian overlay would count as different. The radial key, the
sorted list of distances to the centroid, is unchanged by
rotation and translation, and two keys of equal length are
compared by their mean absolute difference

\begin{equation}
D(a,b)=\frac{1}{N}\sum_{i=1}^{N}\lvert a_{(i)}-b_{(i)}\rvert.
\label{eq:radial}
\end{equation}

The scale \(D_{\mathrm{ave}}\) is the mean pairwise distance of
the first full bank, measured once, so that later members are
judged against a scale the search can no longer move. The merge
distance is then held at

\begin{equation}
D_{\mathrm{merge}}=\tfrac12 D_{\mathrm{ave}}.
\label{eq:merge}
\end{equation}

Conformational space annealing shrinks this cutoff from a wide
value toward a tight one
\cite{leeNewOptimizationMethod1997,leeUnbiasedGlobalOptimization2003}.
Holding it instead keeps a late minimum on the same scale as
the first full bank. An offer inside the merge distance of a
member replaces that member when it is lower. An offer that
matches no member displaces the highest member when it is
lower. A member drawn twenty times without improvement yields
its slot to the next dissimilar fresh start, whatever that
start's energy. The private bank, with one slot per chain and no
splice, is the control for this communication and shares the
same descent.

The bank, the cut-and-splice, and the lattice search are the
methods we extend. The retirement rule, the radial key, the held
merge distance, and the per-chain force ledger keep each shared
minimum tied to the chain that paid for it.

Alongside the bank travel the identities, visit counts, and
well depths of the basins, which are the record the occupancy
law reads. Coordinates change in only two ways: when a chain
adopts a minimum by the rule of Section \ref{sec:flow},
or when the restart below draws a fresh start from a packing
that two peers already share. A descriptor deposit and a gossip
of the well depth update the bias, and every stored structure
is the geometry its own chain quenched.

The two-choice restart reads two peers. When they share a
packing, it broadcasts the fivefold share of that packing and
draws a fresh start from it. Copying coordinates is a separate
step: catalogue adoption, \eqref{eq:adopt}, copies them
only when the parent is a lower minimum of the same packing.

The communicating preset runs one chain on all, single, surface,
shell and sym with Thompson allocation
\cite{thompsonLikelihoodOneUnknown1933}, the return
screen of Section \ref{sec:rscreen}, orbit completion on
each new basin, the symmetrisation of Section
\ref{sec:info}, and recognition of recorded minima. The
bank above is the multi-chain preset.
Below \(98\) atoms the single chain also uses a two-phase surface
move. Hit rates for that preset belong to the campaign that
deposits them.

Algorithm \ref{alg:bank} sets out one trial of the
bank.

\begin{algorithm}[t]
\caption{One trial of the communicating bank, with the radial key of \eqref{eq:radial} and the merge distance of \eqref{eq:merge}.}
\label{alg:bank}
\begin{algorithmic}[1]
\Require \(40\) chains, one slot each, one shared bank
\If{the bank is not full, or a fresh start is drawn}
  \State Draw a random cluster at the density in Table \ref{tbl:params}
\Else
  \State Take the least-used member
  \If{at least two members and a splice is drawn}
    \State Cut-and-splice that member with another
  \Else
    \State Move one to four surface atoms onto vacant hollow sites
  \EndIf
\EndIf
\State Quench with the potential in use
\State Descend at the well of that potential, ranking moves by it
\State Charge every evaluation, including the descent, to this chain
\State Build the radial key \eqref{eq:radial}
\If{the offer falls inside \(D_{\mathrm{merge}}\) and is lower}
  \State Replace that member
\ElsIf{the key matches none and the offer is lower than the highest member}
  \State Displace the highest member
\ElsIf{the offer is fresh and a member has gone \(20\) draws without improvement}
  \State Recycle that slot
\Else
  \State Discard the offer
\EndIf
\end{algorithmic}
\end{algorithm}

The recommended and derived hops are single chains, and
LJ\(_{75}\), the smallest Lennard-Jones cluster whose global minimum
is a Marks decahedron
\cite{marksSurfaceStructureEnergetics1984}, is a hard case
for one. Most of its low
minima are icosahedral, so a chain that has settled among them has
to leave their funnel before it can reach the decahedron
\cite{doyeEvolutionPotentialEnergy1999}. GMIN meets this
cluster with core-orbit symmetrisation
\cite{oakleySymmetrisationSchemesGlobal2013}. Section
\ref{sec:frozen-lj} runs GMIN at that setting against
anneal from the same 64 starts on one host.
\section{Information theory in the hop}
\label{sec:info}
A bank that shares good minima can still hand the next trial to
a move that has stopped finding depth. Four pieces of
information theory guard against that: a posterior on the depth
each move reaches, an entropy that can set the temperature, a
rule for which share is worth broadcasting, and the screen of
\eqref{eq:tau}. The recommended hop also carries one move
taken from geometry rather than information, the symmetrisation of
a cluster onto a point group, and it follows the depth posterior
because that hop is defined there.
\subsection{Depth posterior}

On a deep cluster few of the available moves improve the
incumbent, and an acceptance records only that a trial was kept.
The quantity that should choose the next move is how close to
the incumbent the trial landed, and that depth is the reward the
allocation scores
\cite{thompsonLikelihoodOneUnknown1933},

\begin{equation}
r=-\bigl(E_{\mathrm{new}}-E_{\mathrm{best}}^{+}\bigr)\le 0,
\label{eq:reward}
\end{equation}

where \(E_{\mathrm{best}}^{+}\) is the incumbent after the trial
has been recorded. A trial that sets a new incumbent scores \(0\),
and every other trial scores minus its height above the incumbent.

Each observed depth updates a Normal-Gamma posterior for its
arm, with the right-hand sides evaluated before the update,

\begin{equation}
\begin{aligned}
\kappa&\leftarrow\kappa+1,\\
\mu&\leftarrow(\kappa\mu+r)/(\kappa+1),\\
\alpha&\leftarrow\alpha+1/2,\\
\beta&\leftarrow\beta+\tfrac12\frac{\kappa}{\kappa+1}(r-\mu)^2.
\end{aligned}
\label{eq:ng}
\end{equation}

The update is stationary: every depth an arm has reached counts
with equal weight, and no window or discount forgets an old one.
Each arm starts from \(\mu=0\), \(\kappa=10^{-6}\), \(\alpha=1\) and
\(\beta=1\). Thompson sampling draws once from the predictive distribution of
every arm and plays the largest draw
\cite{thompsonLikelihoodOneUnknown1933}. The next
minimization therefore goes to the move that the posterior, after
its latest update, ranks deepest, and an uncertain arm still wins often enough to
be tested. The predictive draw is Student-\(t\) with \(2\alpha\)
degrees of freedom, centred at \(\mu\), with scale

\begin{equation}
s=\sqrt{\beta(\kappa+1)/(\alpha\kappa)}.
\label{eq:scale}
\end{equation}

The prior is in Table \ref{tbl:params}. The depth
\eqref{eq:reward} is defined on every hop, including a
hop whose quench stopped at the screen, so a trial that saved
its finish still tells the posterior how deep its move went.

A new packing is sometimes one surface atom away and sometimes
several, so one arm makes a random number of surface relocations. The
burst starts with one relocation of a surface atom and continues
with probability \(2/3\), up to eight relocations,

\begin{equation}
P(K=k)=\tfrac13\bigl(\tfrac23\bigr)^{k-1},
\quad k=1,\ldots,7,
\qquad
P(K=8)=\bigl(\tfrac23\bigr)^{7}.
\label{eq:burst}
\end{equation}

A single acceptance test is applied to the endpoint of the
burst, so a chain can cross several surface sites before it is
judged. The uncapped geometric law has mean \(3\); the mean under
the cap is in the Supporting Information. Figure
\ref{fig:rec-arms} shows the arms among which this
draw is made.

The recommended hop is the cluster preset with this burst
library and the depth reward, plus the trail, the climb, and the
restart, which act when the search stalls, and the symmetrisation
set out below. It holds the cluster
temperature \(T=0.8 E_0\) and leaves the statistical temperature
and the budget window unused.

The driver accepts on the quenched energy plus a well-tempered
bias. Barducci, Bussi, and Parrinello designed that deposit so
that a frequently visited collective variable fills in while
the rest of the surface stays open
\cite{barducciWelltemperedMetadynamicsSmoothly2008}.
The deposit at a collective variable \(s_k\) is

\begin{equation}
w_k=w_0\exp\bigl(-V_{k-1}(s_k)/((\gamma-1)T)\bigr).
\label{eq:bias}
\end{equation}

The tested difference is

\begin{equation}
\Delta=\bigl(E_q(y)+V(s_y)\bigr)-\bigl(E_q(x)+V(s_x)\bigr),
\label{eq:biased-metro}
\end{equation}

and the chain accepts when \(\Delta<0\) or when a uniform draw
falls below \(\exp(-\Delta/T)\). With the bias at zero,
\eqref{eq:biased-metro} is the basin-hopping test of
Section \ref{sec:background}, so the unbiased hop
accepts exactly as basin hopping does. The height \(w_0\) and the
factor \(\gamma\) are in Table \ref{tbl:params}. A
screened trial enters \eqref{eq:reward} and
\eqref{eq:biased-metro} with its screened energy, and
the catalogue records a minimum only after a finished quench.

\begin{figure}[t]
\centering
\begin{tikzpicture}[
  every node/.style={font=\sffamily\scriptsize, align=center},
  arm/.style={draw=RuhiTeal, fill=white, rounded corners=2pt,
              minimum width=1.45cm, minimum height=0.55cm,
              line width=0.7pt, inner sep=1pt},
  box/.style={arm, minimum width=2.4cm, minimum height=0.62cm},
  arrow/.style={-{Stealth[length=4.5pt]}, RuhiTeal, line width=0.55pt}
]
  \node[arm] (a) at (0,0) {single};
  \node[arm] (b) at (1.7,0) {surface};
  \node[arm] (c) at (3.4,0) {sym};
  \node[arm] (d) at (5.1,0) {burst};
  \node[arm] (e) at (6.8,0) {soap};
  \node[box] (q) at (3.4,-1.45) {one quench};
  \node[box, fill=RuhiYellow] (met) at (3.4,-2.7) {one acceptance test};
  \draw[arrow] (a.south) -- (a.south |- q.north);
  \draw[arrow] (b.south) -- (b.south |- q.north);
  \draw[arrow] (c) -- (q);
  \draw[arrow] (d.south) -- (d.south |- q.north);
  \draw[arrow] (e.south) -- (e.south |- q.north);
  \draw[arrow] (q) -- (met);
\end{tikzpicture}
\caption{The arms of the recommended hop. Each proposal is quenched once and meets one acceptance test. The burst length follows \eqref{eq:burst}.}
\label{fig:rec-arms}
\end{figure}
\subsection{Move library}
\label{sec:arms}
Every arm acts on the positions \(x_a\in\mathbb{R}^3\),
\(a=1,\ldots,n\), in units of \(\sigma\), with
\(U\sim\mathcal{U}[0,1)\) drawn afresh at each use, \(c\) the
centroid and \(\hat u\) a uniform unit vector. The single arm
displaces one atom \(i\) drawn uniformly,

\begin{equation}
x'_{i,k}=x_{i,k}+\xi_k,\qquad \xi_k\sim\mathcal{U}[-h,h),\quad k=1,2,3,\qquad h=1.0.
\label{eq:arm-single}
\end{equation}

The surface arm moves the least coordinated atom
\(w=\arg\min_a z_a\), with
\(z_a=\#\{b\neq a:\lVert x_a-x_b\rVert<1.6\}\) and the lowest index
taken on a tie, to the rim of the rest of the cluster,

\begin{equation}
x'_w=c+\rho\,(0.85+0.20\,U)\,\hat u,
\qquad
\rho=\max_{a\neq w}\lVert x_a-c\rVert.
\label{eq:arm-surface}
\end{equation}

The sym arm draws an axis \(\hat e\), an order
\(m\sim\mathcal{U}\{2,\ldots,6\}\) and a blend
\(\beta=0.3+0.4\,U\). With \(v_a=x_a-c\),
\(w_b=R(\hat e,2\pi/m)\,v_b\) and
\(b^*(a)=\arg\min_b\lVert v_a-w_b\rVert\), every atom moves from the
same \(x\) as

\begin{equation}
x'_a=
\begin{cases}
c+(1-\beta)\,v_a+\beta\,w_{b^*(a)}, & \lVert v_a-w_{b^*(a)}\rVert<2.5,\\
x_a, & \text{otherwise}.
\end{cases}
\label{eq:arm-sym}
\end{equation}

The burst arm applies \eqref{eq:arm-surface} \(K\) times, with \(K\)
drawn from \eqref{eq:burst} and the coordination, centroid
and rim recomputed before each relocation. On a Lennard-Jones
cluster the soap arm is a fivefold hop. Its candidate axes are the
six icosahedral axes \((0,\pm1,\varphi)\) and the directions from
\(c\) to the 12 atoms farthest from it, with any axis inside
\(\lvert\hat a\cdot\hat a'\rvert>0.97\) of a kept one dropped. About
an axis \(\hat a\), with \(R_5=R(\hat a,2\pi/5)\) and \(\pi\) the
assignment that minimises the sum,

\begin{equation}
\delta r_i=x_i-(R_5x)_{\pi(i)},\qquad
d_5(\hat a)=\Bigl(\tfrac1n\textstyle\sum_i\lVert\delta r_i\rVert^2\Bigr)^{1/2}.
\label{eq:arm-d5}
\end{equation}

For \(n>13\) the arm keeps the axes with \(d_5<1.40\), needs at
least two, and takes with equal probability the one whose \(d_5\)
lies nearest \(0.90\) or nearest \(0.99\). It then moves the five
atoms of largest \(\lVert\delta r_i\rVert\),

\begin{equation}
x'_i=x_i+0.75\,\frac{\delta r_i}{\bigl(\tfrac15\sum_{j\in\mathcal{S}}\lVert\delta r_j\rVert^2\bigr)^{1/2}},\qquad i\in\mathcal{S},\ \lvert\mathcal{S}\rvert=5.
\label{eq:arm-soap}
\end{equation}

With fewer than two axes it falls back to the best axis when
\(\min d_5\le1.15\). An arm that leaves \(x\) unchanged is redrawn
from the others. Two further arms act on the whole cluster. The all
arm displaces every coordinate by \(\mathcal{U}[-0.38,0.38)\), and
the shell arm rotates the atoms beyond the median radius by
\(0.3+(\pi-0.3)\,U\) about \(\hat u\). After each proposal the
cluster is recentred and held inside the radius \(0.9\sqrt[3]{n}\).

Figure \ref{fig:rec-arms} is the library of the
recommended hop for \(n\neq38\). Every campaign of Section
\ref{sec:results} draws instead from all, single,
surface, shell and sym. The frozen LJ\(_{38}\) and LJ\(_{75}\) hop
allocates among them by \eqref{eq:reward} and
\eqref{eq:ng}, and the ABCluster and Morse campaigns draw
them uniformly.
\subsection{Return screen}
\label{sec:rscreen}
A trial that falls back into the basin it left costs a full
quench and finds nothing. After the \(S=25\)-step screen, the
return screen compares the sorted vector \(D(x)\) of all
\(n(n-1)/2\) pair distances at the screened point and at the
current minimum,

\begin{equation}
\Delta_D=\bigl\lVert D(x_{\mathrm{sc}})-D(x)\bigr\rVert_2,
\qquad
\mathrm{rscreen}\iff \Delta_D<r_m=0.7.
\label{eq:rscreen}
\end{equation}

A returning trial keeps its screened energy and skips the rest of
the quench. It still meets \eqref{eq:biased-metro} and
deposits a bias, but it is never recorded, never moves
\(E_{\mathrm{best}}\), and never enters the predictive screen. The
radius \(r_m\) is the one that merges two minima into one basin.
Under the recognition policy of the communicating preset, a
screened point within \(r_m\) of any of the last 512 recorded
minima takes that minimum's energy and coordinates and also skips
the quench.
\subsection{Symmetrisation}

The global minimum of a cluster is often more symmetric than the
minima around it, and a chain that relocates a few surface atoms at
a time can pass close to such a structure without reaching it. On
LJ\(_{75}\) the global minimum is a Marks decahedron of point group
D\(_{5~h}\) \cite{marksSurfaceStructureEnergetics1984}, and
most of the low minima around it are icosahedral
\cite{doyeEvolutionPotentialEnergy1999}. Oakley, Johnston,
and Wales symmetrise the structure itself in GMIN, either by finding
a symmetric core and filling the incomplete orbits outside it, or
by averaging the whole structure over a point group under a
continuous symmetry measure
\cite{oakleySymmetrisationSchemesGlobal2013}. The hop takes
both into the chain.

For the average, the cluster is centred on its centroid, scaled so
that its farthest atom lies at unit radius, and turned to an
orientation, which gives the points \(y_i\). Each operation \(g\)
of a point group \(G\) is matched to the structure by the assignment
\(p_g\) of atoms of the same species that minimizes
\(\sum_i\lVert y_i-g\,y_{p_g(i)}\rVert^2\), and the average and its
measure are

\begin{equation}
\hat y_i=\frac{1}{|G|}\sum_{g\in G}g\,y_{p_g(i)},
\qquad
S=\frac{\sum_i\lVert y_i-\hat y_i\rVert^2}{\sum_i\lVert y_i\rVert^2}.
\label{eq:csm}
\end{equation}

A structure that has the symmetry of \(G\) in that orientation has
\(S=0\), and a larger \(S\) means that the average moves its atoms
further. Every tenth quench the hop forms this average for
C\(_i\), C\(_5\), and D\(_{5~h}\) in six orientations, each principal
axis of the inertia tensor taken as the symmetry axis with and
without a fixed fifth of a turn about it. It keeps the group and the
orientation of smallest \(S\), undoes the turn and the scaling, and
quenches the result, and the chain adopts that minimum only when it
lies below the current one. An average that brings two atoms closer
than the pair limit of Table \ref{tbl:params} is dropped,
because a quench started there begins at a near-singular force. The
average draws no random number, so it leaves the stream of the
chain where the seed put it.

The core orbits act when the chain has stalled among icosahedral
minima. When the common-neighbour signature (555) of the icosahedron
marks more than the plateau fraction of Table
\ref{tbl:params} of the neighbour pairs, and five hops have
passed without a new incumbent, the chain draws a fresh random
cluster from its own stream, fills the incomplete orbits of that
cluster's symmetric core, averages it by
\eqref{eq:csm}, quenches it, and adopts the result even
when it lies higher. Five hops is the short reseeding interval that
Oakley, Johnston, and Wales found to pay on LJ\(_{75}\)
\cite{oakleySymmetrisationSchemesGlobal2013}. The
recommended and communicating presets carry both parts, the derived
preset inherits them from the recommended hop, and the cluster
preset leaves them off. A chart of rigid molecules turns the move
off, because its coordinates are centres and rotations rather than
atoms.
\subsection{Entropy}

A held temperature ignores what the census has learned. As the
occupied quenched states fill in, their entropy \(S(E)\) measures
how much new packing a further drop in energy buys, and its
slope defines a temperature. When the entropy estimator is on,
the statistical temperature is the reciprocal of that slope,

\begin{equation}
T=\mathrm{clamp}\bigl(1/S'(E),\; 0.2\,T_{\mathrm{cfg}},\; 5\,T_{\mathrm{cfg}}\bigr).
\label{eq:entropy}
\end{equation}

The slope is a central difference of the fitted density of
states. A finite positive temperature from that slope is clamped
to the band from \(0.2\,T_{\mathrm{cfg}}\) to
\(5\,T_{\mathrm{cfg}}\) around the configured temperature, and
any other result leaves the current temperature in place. On a
ladder, only the rung with temperature ratio \(1\) writes into
the density while the estimator is on. The recommended hop holds
\(T=0.8 E_0\) and leaves the estimator unused.
\subsection{Channel value}

A restart should broadcast the more informative of two shares,
but with counts from only two peers a ratio of
frequencies is unstable. Two binary broadcasts can instead be
compared by a cross-multiplication that stays in the integers.
For crossing counts \(f_c,f_c'\) and the remaining counts
\(f_f,f_f'\), the hypotheses \(f_c' f_f\le f_c f_f'\) and
\(f_c' f_c\le f_c f_c'\) imply

\begin{equation}
f_c'(f_c+f_f)\le f_c(f_c'+f_f').
\label{eq:sharper}
\end{equation}

The fivefold detector applies \eqref{eq:sharper} to the
counts \((2,3)\) against \((2,500)\), whose posterior shares are
\(2/5\) and \(2/502\). Sending the smaller posterior to \(n\)
receivers multiplies its value by \(n\), while a single copy is
worth only the share itself. The restart of Section
\ref{sec:chains} broadcasts the fivefold share.
\subsection{Posterior screen}

The screen and the posterior act together within one hop. The
threshold is \eqref{eq:tau}, and the probability it is
compared with is the Student-\(t\) tail of Section
\ref{sec:flow}. Algorithm \ref{alg:thompson}
sets out the whole hop.

\begin{algorithm}[t]
\caption{One hop under the depth posterior. The update is \eqref{eq:ng}. The burst is \eqref{eq:burst}. The screen is \eqref{eq:tau}. The symmetrised average is \eqref{eq:csm}.}
\label{alg:thompson}
\begin{algorithmic}[1]
\State Draw an arm, including the population-annealing resample
\State Propose. A burst has length \eqref{eq:burst} and returns one endpoint
\State Relax \(S=25\) steps and read the screened energy
\If{the return screen \eqref{eq:rscreen} fires}
  \State Keep the screened energy and skip the quench
\ElsIf{the Student-\(t\) screen is on}
  \State Finish the quench unless the screened energy is worse and the predictive tail falls below \(\tau\)
\Else
  \State Finish the quench unless \eqref{eq:screen-margin} holds
\EndIf
\State Accept or reject by \eqref{eq:biased-metro}
\State Observe the depth \eqref{eq:reward} and apply \eqref{eq:ng}
\State Update \(n\) and \(n_1\), and test the hatch
\State On a stall, update the trail, the climb, or the restart by \eqref{eq:ng}
\If{this is a tenth quench}
  \State Quench the average \eqref{eq:csm} of least \(S\) over C\(_i\), C\(_5\), D\(_{5~h}\), and adopt it if lower
\EndIf
\If{five hops on an icosahedral plateau have found no new incumbent}
  \State Fill the core orbits of a fresh cluster, average it by \eqref{eq:csm}, quench, and adopt
\EndIf
\end{algorithmic}
\end{algorithm}
\section{Parameters}
\label{sec:params}
Table \ref{tbl:params} lists every constant used by the
equations and the three algorithms. The seeds and the force caps belong to the
campaign design, and the parameters of each potential are fixed
in the build before the first seed.

\begin{table}[t]
\centering
\scriptsize
\setlength{\tabcolsep}{3pt}
\caption{Constants of the hop and the bank, in reduced units.}
\label{tbl:params}
\begin{adjustbox}{max width=\textwidth}
\begin{tabular}{@{}>{\raggedright\arraybackslash}p{2.0cm}>{\raggedright\arraybackslash}p{3.5cm}>{\raggedright\arraybackslash}p{4.6cm}>{\raggedright\arraybackslash}p{4.2cm}@{}}
\toprule
Symbol & Meaning & Config key & Value\\
\midrule
\(T\) & basin-hopping temperature & \texttt{temperature} & \(0.8 E_0\)\\
\(\theta\) & window design point & \texttt{theta} & \(1/2\)\\
\(S,R\) & screen steps, relax steps & \texttt{screen\_steps}, \texttt{relax\_steps} & \(25\), \(200\)\\
margin & screen margin \eqref{eq:screen-margin} & \texttt{screen\_margin} & \(2 E_0\)\\
\(A\) & margin allowance: quantile, window, warmup, audit & \texttt{calibrated\_screen} & \(0.9\), \(256\), \(8\), every \(8\)th\\
\(r_m\) & return radius \eqref{eq:rscreen}, basin merge & \texttt{merge\_radius} & \(0.7\,\sigma\)\\
\(\tau\) & screen threshold & \texttt{bayes\_threshold} & \(7/15\), screen on\\
exploration & screen exploration floor & \texttt{bayes\_exploration} & \(0.1\)\\
warmup & observations before the posterior decides & \texttt{bayes\_warmup} & \(300\)\\
\(\alpha\) & occupancy ceiling & census constants & \(1/5\)\\
\(k,n_{\min}\) & singleton budget, visit floor & \texttt{gt\_min\_visits} & \(3\), \(20\)\\
\(\gamma,w_0\) & bias factor and height & \texttt{bias\_gamma}, \texttt{bias\_height} & \(5\), \(0.25 E_0\)\\
\(P(K=k)\) & burst continuation & LeanBurst & \eqref{eq:burst}\\
prior & depth posterior \(\kappa_0,\alpha_0,\beta_0,\mu_0\) & \texttt{DepthAllocator} & \(10^{-6}\), \(1\), \(1\), \(0\)\\
regressors & screen design vector & partial relaxation & \((1,E_{\mathrm{sc}},\Delta_D,E_{\mathrm{sc}}\Delta_D)\)\\
screen prior & precision, shape, rate & posterior screen & \(10^{-3}I\), \(10^{-3}\), \(10^{-3}\)\\
weights & energy, novelty, scarcity, uncertainty & \texttt{SelectionCoefficients} & \(1\), \(0.8\), \(0.6\), \(0.4\)\\
clip & log-weight floor & \texttt{log\_weight\_clip} & \(4\)\\
cap & offspring cap & \(\lceil R/2\rceil\) & \(\lceil R/2\rceil\)\\
\(\sigma_t\) & adoption jitter & \texttt{CATALOG\_\allowbreak{}TRANSPORT\_\allowbreak{}NOISE} & \(0.05 L_0\)\\
undercut & adoption and replacement gaps & energy tests & \(10^{-10}\), \(10^{-7}\)\\
chains, slots & bank width & \texttt{Plan} & \(40\), \(1\)\\
fresh, splice & draw probabilities & \texttt{Plan} & \(0.3\), \(0.2\)\\
moved & surface atoms & \texttt{Plan.moved} & \(1\) to \(4\)\\
merge & held fractions & \texttt{Plan.merge} & \(0.5\), \(0.5\)\\
retire & draws without improvement & \texttt{Plan.retire} & \(20\)\\
density & random-start density & \texttt{Plan.density} & \(0.7\)\\
separation & closest approach & \texttt{Plan.min\_separation} & \(0.85\)\\
\(q_v,q_a\) & GSA defaults & Python \texttt{Gsa} & \(2.62\), \(1.7\)\\
\(k_0\) & Boltzmann offset & \texttt{LogCool} & \(2\)\\
stall & trail, climb, restart & recommended hop & all three on\\
groups & symmetrised average & \texttt{csm::propose} & C\(_i\), C\(_5\), D\(_{5~h}\), six orientations\\
interval, reseed & quenches between averages, stalled hops before a reseed & \texttt{ContinuousSymmetry::\allowbreak{}Measure} & \(10\), \(5\)\\
pair limit & closest pair of an average & \texttt{csm::SINGULAR\_PAIR} & \(10^{-2}\)\\
plateau & (555) fraction of the neighbour pairs, cutoff & \texttt{ih\_dominated} & \(0.12\), \(1.35\)\\
screen & posterior predictive & cluster preset & Student-\(t\), on\\
\bottomrule
\end{tabular}
\end{adjustbox}
\end{table}
\section{Results}
\label{sec:results}
Three campaigns give the results, and each fixes its starts, its
budget and its host before the first seed. Section
\ref{sec:frozen-lj} runs anneal, GMIN, ABCluster and plain
basin hopping from 64 shared OptBench starts of LJ\(_{38}\) and
LJ\(_{75}\). Section \ref{sec:abcluster} runs anneal
against ABCluster on LJ\(_{55}\), LJ\(_{98}\) and Morse clusters
from 32 paired seeds. and Section
\ref{sec:morse-gmin} runs anneal against GMIN on the
Cambridge Morse table. Figure
\ref{fig:summary} collects the hit rates of all three. The
Supporting Information gives the ablation of the components of the
hop, every Morse target and the command of each campaign.
\subsection{Computational details}
\label{sec:compdet}
The Lennard-Jones and Morse
\cite{morseDiatomicMoleculesAccording1929} surfaces of
the three campaigns are those anneal builds in, the reduced
pair \(4(r^{-12}-r^{-6})\) and the Morse pair
\(e^{\rho(1-r)}(e^{\rho(1-r)}-2)\), neither with a cutoff. GMIN and ABCluster run their own forms of the same
surfaces.
\subsubsection{Hardware}
\label{sec:hardware}
Each campaign ran all its codes on one host, one thread per run.
The LJ\(_{38}\) and LJ\(_{75}\) campaign ran on 8 Intel Xeon cores
under Linux 6.12.94, with anneal built by rustc 1.88.0 and GMIN by
gfortran 13.3.0. The ABCluster campaign ran on an Intel Xeon at
2400 MHz with 8 CPUs, with anneal built by rustc 1.91.0 and
ABCluster by g++ 13.3.0. The Morse campaign ran on an Intel Xeon
family 6 model 207, stepping 2, with 8 cores at 1 thread each
under KVM, with anneal built by rustc 1.99.1 and GMIN by gfortran
13.3.0. Wall times compare codes only within a campaign.
\subsection{Metrics}
\label{sec:metrics}
A run hits when a minimum it reaches lies below the reference
energy plus a tolerance, 0.001 on LJ\(_{38}\) and LJ\(_{75}\),
\(10^{-4}\) on the Cambridge Morse table, and the value the
ABCluster protocol fixes for each of its systems. For \(k\) hits
in \(n\) seeds the rate \(\hat p=k/n\) carries the Wilson score
interval \cite{wilsonProbableInferenceLaw1927}

\begin{equation}
\frac{\hat p+\frac{z^2}{2n}\pm z\sqrt{\frac{\hat p(1-\hat p)}{n}+\frac{z^2}{4n^2}}}{1+\frac{z^2}{n}},
\qquad z=1.96,
\label{eq:wilson}
\end{equation}

The interval stays inside \([0,1]\) when a code hits on every
seed or on none. Cost is counted in force calls, each code by its
own counter, and summarized by the penalized cost PAR-2,

\begin{equation}
\mathrm{PAR2}_i=
\begin{cases}
c_i, & \text{seed } i \text{ hits},\\
2\,C, & \text{otherwise},
\end{cases}
\label{eq:par2}
\end{equation}

where \(c_i\) is the force calls of seed \(i\) to its first hit
and \(C\) the budget of a seed, so a miss is counted and pays twice
the budget. Each table states whether it gives the median or the
mean of \(\mathrm{PAR2}_i\) over its seeds.
\subsection{Across the campaigns}
\label{sec:summary}
Figure \ref{fig:summary} shades the hit rate of every code
on every system of the three campaigns. A crossed cell is a code the
campaign did not run on that system.

\begin{figure}[t]
\centering
\begin{tikzpicture}[x=1.45cm,y=0.36cm,font=\scriptsize]
\node[anchor=south,inner sep=2pt] at (0.5,0) {anneal};
\node[anchor=south,inner sep=2pt] at (1.5,0) {plain BH};
\node[anchor=south,inner sep=2pt] at (2.5,0) {GMIN};
\node[anchor=south,inner sep=2pt] at (3.5,0) {ABCluster};
\node[anchor=east,inner sep=3pt] at (0,-0.50) {LJ\(_{38}\)};
\draw[draw=RuhiTeal!35,fill=RuhiTeal!100] (0,-1.00) rectangle (1,0.00);
\node[font=\tiny,text=white] at (0.5,-0.50) {64/64};
\draw[draw=RuhiTeal!35,fill=RuhiTeal!97] (1,-1.00) rectangle (2,0.00);
\node[font=\tiny,text=white] at (1.5,-0.50) {62/64};
\draw[draw=RuhiTeal!35,fill=RuhiTeal!100] (2,-1.00) rectangle (3,0.00);
\node[font=\tiny,text=white] at (2.5,-0.50) {64/64};
\draw[draw=RuhiTeal!35,fill=RuhiTeal!100] (3,-1.00) rectangle (4,0.00);
\node[font=\tiny,text=white] at (3.5,-0.50) {64/64};
\node[anchor=east,inner sep=3pt] at (0,-1.50) {LJ\(_{75}\)};
\draw[draw=RuhiTeal!35,fill=RuhiTeal!100] (0,-2.00) rectangle (1,-1.00);
\node[font=\tiny,text=white] at (0.5,-1.50) {64/64};
\draw[draw=RuhiTeal!35,fill=RuhiTeal!2] (1,-2.00) rectangle (2,-1.00);
\node[font=\tiny,text=black] at (1.5,-1.50) {1/64};
\draw[draw=RuhiTeal!35,fill=RuhiTeal!100] (2,-2.00) rectangle (3,-1.00);
\node[font=\tiny,text=white] at (2.5,-1.50) {64/64};
\draw[draw=RuhiTeal!35,fill=white] (3,-2.00) rectangle (4,-1.00);
\node[font=\tiny] at (3.5,-1.50) {0/6 finished};
\node[anchor=east,inner sep=3pt] at (0,-2.90) {LJ\(_{55}\)};
\draw[draw=RuhiTeal!35,fill=RuhiTeal!100] (0,-3.40) rectangle (1,-2.40);
\node[font=\tiny,text=white] at (0.5,-2.90) {32/32};
\draw[draw=RuhiTeal!35,fill=white] (1,-3.40) rectangle (2,-2.40);
\draw[gray!60] (1.1,-3.30) -- (1.9,-2.50) (1.1,-2.50) -- (1.9,-3.30);
\draw[draw=RuhiTeal!35,fill=white] (2,-3.40) rectangle (3,-2.40);
\draw[gray!60] (2.1,-3.30) -- (2.9,-2.50) (2.1,-2.50) -- (2.9,-3.30);
\draw[draw=RuhiTeal!35,fill=RuhiTeal!100] (3,-3.40) rectangle (4,-2.40);
\node[font=\tiny,text=white] at (3.5,-2.90) {32/32};
\node[anchor=east,inner sep=3pt] at (0,-3.90) {LJ\(_{98}\)};
\draw[draw=RuhiTeal!35,fill=RuhiTeal!6] (0,-4.40) rectangle (1,-3.40);
\node[font=\tiny,text=black] at (0.5,-3.90) {2/32};
\draw[draw=RuhiTeal!35,fill=white] (1,-4.40) rectangle (2,-3.40);
\draw[gray!60] (1.1,-4.30) -- (1.9,-3.50) (1.1,-3.50) -- (1.9,-4.30);
\draw[draw=RuhiTeal!35,fill=white] (2,-4.40) rectangle (3,-3.40);
\draw[gray!60] (2.1,-4.30) -- (2.9,-3.50) (2.1,-3.50) -- (2.9,-4.30);
\draw[draw=RuhiTeal!35,fill=RuhiTeal!3] (3,-4.40) rectangle (4,-3.40);
\node[font=\tiny,text=black] at (3.5,-3.90) {1/32};
\node[anchor=east,inner sep=3pt] at (0,-4.90) {M\(_{55}\), \(\rho=6\)};
\draw[draw=RuhiTeal!35,fill=RuhiTeal!100] (0,-5.40) rectangle (1,-4.40);
\node[font=\tiny,text=white] at (0.5,-4.90) {32/32};
\draw[draw=RuhiTeal!35,fill=white] (1,-5.40) rectangle (2,-4.40);
\draw[gray!60] (1.1,-5.30) -- (1.9,-4.50) (1.1,-4.50) -- (1.9,-5.30);
\draw[draw=RuhiTeal!35,fill=white] (2,-5.40) rectangle (3,-4.40);
\draw[gray!60] (2.1,-5.30) -- (2.9,-4.50) (2.1,-4.50) -- (2.9,-5.30);
\draw[draw=RuhiTeal!35,fill=RuhiTeal!22] (3,-5.40) rectangle (4,-4.40);
\node[font=\tiny,text=black] at (3.5,-4.90) {7/32};
\node[anchor=east,inner sep=3pt] at (0,-5.90) {M\(_{55}\), \(\rho=14\)};
\draw[draw=RuhiTeal!35,fill=RuhiTeal!0] (0,-6.40) rectangle (1,-5.40);
\node[font=\tiny,text=black] at (0.5,-5.90) {0/32};
\draw[draw=RuhiTeal!35,fill=white] (1,-6.40) rectangle (2,-5.40);
\draw[gray!60] (1.1,-6.30) -- (1.9,-5.50) (1.1,-5.50) -- (1.9,-6.30);
\draw[draw=RuhiTeal!35,fill=white] (2,-6.40) rectangle (3,-5.40);
\draw[gray!60] (2.1,-6.30) -- (2.9,-5.50) (2.1,-5.50) -- (2.9,-6.30);
\draw[draw=RuhiTeal!35,fill=RuhiTeal!0] (3,-6.40) rectangle (4,-5.40);
\node[font=\tiny,text=black] at (3.5,-5.90) {0/32};
\node[anchor=east,inner sep=3pt] at (0,-6.90) {M\(_{80}\), \(\rho=6\)};
\draw[draw=RuhiTeal!35,fill=RuhiTeal!19] (0,-7.40) rectangle (1,-6.40);
\node[font=\tiny,text=black] at (0.5,-6.90) {6/32};
\draw[draw=RuhiTeal!35,fill=white] (1,-7.40) rectangle (2,-6.40);
\draw[gray!60] (1.1,-7.30) -- (1.9,-6.50) (1.1,-6.50) -- (1.9,-7.30);
\draw[draw=RuhiTeal!35,fill=white] (2,-7.40) rectangle (3,-6.40);
\draw[gray!60] (2.1,-7.30) -- (2.9,-6.50) (2.1,-6.50) -- (2.9,-7.30);
\draw[draw=RuhiTeal!35,fill=RuhiTeal!0] (3,-7.40) rectangle (4,-6.40);
\node[font=\tiny,text=black] at (3.5,-6.90) {0/32};
\node[anchor=east,inner sep=3pt] at (0,-7.90) {M\(_{80}\), \(\rho=14\)};
\draw[draw=RuhiTeal!35,fill=RuhiTeal!31] (0,-8.40) rectangle (1,-7.40);
\node[font=\tiny,text=black] at (0.5,-7.90) {10/32};
\draw[draw=RuhiTeal!35,fill=white] (1,-8.40) rectangle (2,-7.40);
\draw[gray!60] (1.1,-8.30) -- (1.9,-7.50) (1.1,-7.50) -- (1.9,-8.30);
\draw[draw=RuhiTeal!35,fill=white] (2,-8.40) rectangle (3,-7.40);
\draw[gray!60] (2.1,-8.30) -- (2.9,-7.50) (2.1,-7.50) -- (2.9,-8.30);
\draw[draw=RuhiTeal!35,fill=RuhiTeal!0] (3,-8.40) rectangle (4,-7.40);
\node[font=\tiny,text=black] at (3.5,-7.90) {0/32};
\node[anchor=east,inner sep=3pt] at (0,-9.30) {Cambridge Morse, \(\rho=3\)};
\draw[draw=RuhiTeal!35,fill=RuhiTeal!81] (0,-9.80) rectangle (1,-8.80);
\node[font=\tiny,text=white] at (0.5,-9.30) {791/976};
\draw[draw=RuhiTeal!35,fill=white] (1,-9.80) rectangle (2,-8.80);
\draw[gray!60] (1.1,-9.70) -- (1.9,-8.90) (1.1,-8.90) -- (1.9,-9.70);
\draw[draw=RuhiTeal!35,fill=RuhiTeal!71] (2,-9.80) rectangle (3,-8.80);
\node[font=\tiny,text=white] at (2.5,-9.30) {693/976};
\draw[draw=RuhiTeal!35,fill=white] (3,-9.80) rectangle (4,-8.80);
\draw[gray!60] (3.1,-9.70) -- (3.9,-8.90) (3.1,-8.90) -- (3.9,-9.70);
\node[anchor=east,inner sep=3pt] at (0,-10.30) {Cambridge Morse, \(\rho=6\)};
\draw[draw=RuhiTeal!35,fill=RuhiTeal!39] (0,-10.80) rectangle (1,-9.80);
\node[font=\tiny,text=black] at (0.5,-10.30) {378/976};
\draw[draw=RuhiTeal!35,fill=white] (1,-10.80) rectangle (2,-9.80);
\draw[gray!60] (1.1,-10.70) -- (1.9,-9.90) (1.1,-9.90) -- (1.9,-10.70);
\draw[draw=RuhiTeal!35,fill=RuhiTeal!72] (2,-10.80) rectangle (3,-9.80);
\node[font=\tiny,text=white] at (2.5,-10.30) {704/976};
\draw[draw=RuhiTeal!35,fill=white] (3,-10.80) rectangle (4,-9.80);
\draw[gray!60] (3.1,-10.70) -- (3.9,-9.90) (3.1,-9.90) -- (3.9,-10.70);
\node[anchor=east,inner sep=3pt] at (0,-11.30) {Cambridge Morse, \(\rho=10\)};
\draw[draw=RuhiTeal!35,fill=RuhiTeal!10] (0,-11.80) rectangle (1,-10.80);
\node[font=\tiny,text=black] at (0.5,-11.30) {101/976};
\draw[draw=RuhiTeal!35,fill=white] (1,-11.80) rectangle (2,-10.80);
\draw[gray!60] (1.1,-11.70) -- (1.9,-10.90) (1.1,-10.90) -- (1.9,-11.70);
\draw[draw=RuhiTeal!35,fill=RuhiTeal!4] (2,-11.80) rectangle (3,-10.80);
\node[font=\tiny,text=black] at (2.5,-11.30) {35/976};
\draw[draw=RuhiTeal!35,fill=white] (3,-11.80) rectangle (4,-10.80);
\draw[gray!60] (3.1,-11.70) -- (3.9,-10.90) (3.1,-10.90) -- (3.9,-11.70);
\node[anchor=east,inner sep=3pt] at (0,-12.30) {Cambridge Morse, \(\rho=14\)};
\draw[draw=RuhiTeal!35,fill=RuhiTeal!5] (0,-12.80) rectangle (1,-11.80);
\node[font=\tiny,text=black] at (0.5,-12.30) {51/976};
\draw[draw=RuhiTeal!35,fill=white] (1,-12.80) rectangle (2,-11.80);
\draw[gray!60] (1.1,-12.70) -- (1.9,-11.90) (1.1,-11.90) -- (1.9,-12.70);
\draw[draw=RuhiTeal!35,fill=RuhiTeal!1] (2,-12.80) rectangle (3,-11.80);
\node[font=\tiny,text=black] at (2.5,-12.30) {7/976};
\draw[draw=RuhiTeal!35,fill=white] (3,-12.80) rectangle (4,-11.80);
\draw[gray!60] (3.1,-12.70) -- (3.9,-11.90) (3.1,-11.90) -- (3.9,-12.70);
\end{tikzpicture}
\caption{Seeds that hit, of the seeds run, for every code on every system of the three campaigns, darker for a higher fraction. From the top, the rows are LJ\(_{38}\) and LJ\(_{75}\) from 64 shared starts (Table~\ref{tbl:frozen-lj}), the systems against ABCluster (Table~\ref{tbl:abcluster}), and the Cambridge Morse table against GMIN, pooled over the 61 targets of each range (Table~\ref{tbl:morse-rho}). A crossed cell is a code the campaign did not run. On LJ\(_{75}\) six ABCluster seeds had finished, none of them hitting, and the other 58 were still running, so that cell is left white. The anneal column is the configuration of each campaign, given in its table.}
\label{fig:summary}
\end{figure}
\subsection{LJ38 and LJ75 on shared starts}
\label{sec:frozen-lj}
\begin{figure}[t]
\centering
\includegraphics[width=0.49\textwidth]{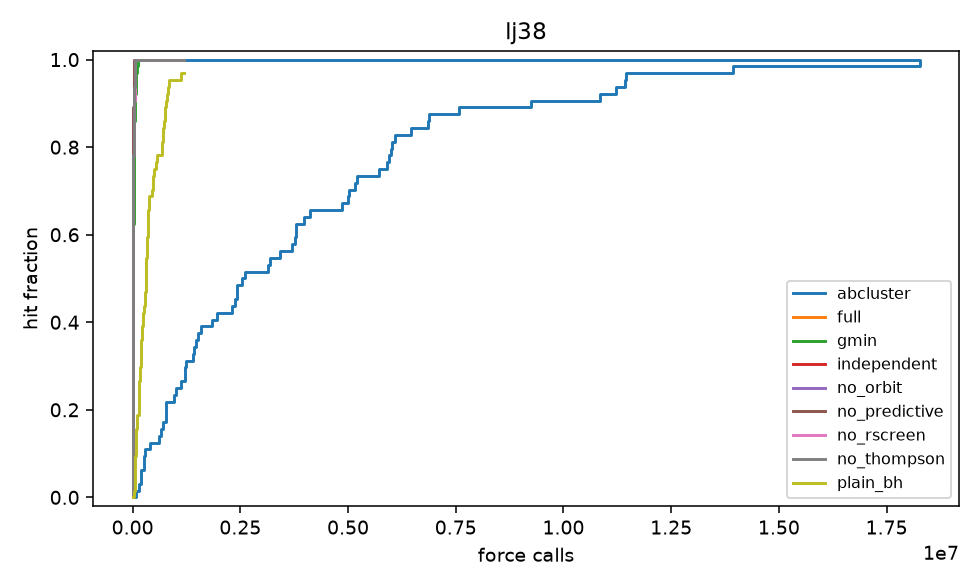}\hfill
\includegraphics[width=0.49\textwidth]{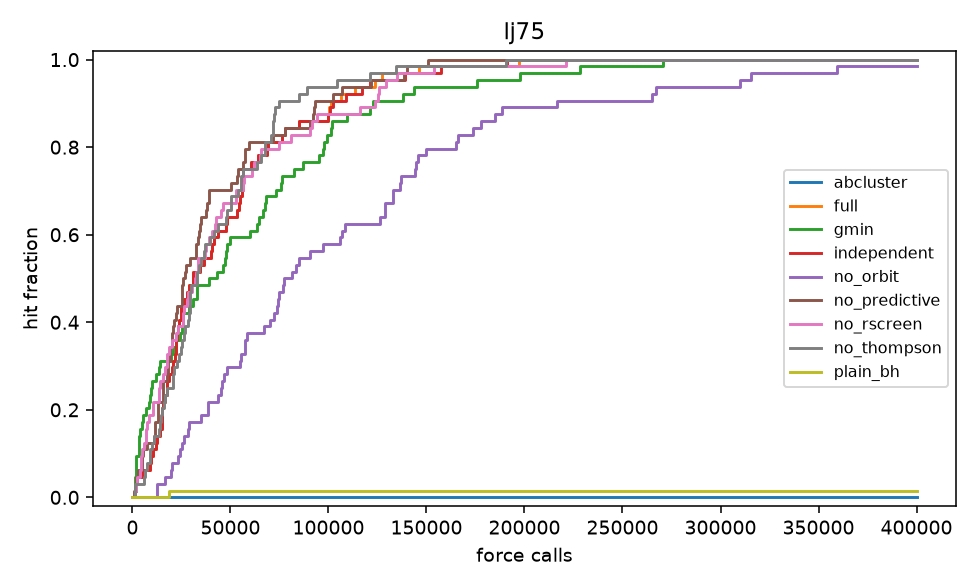}
\caption{Fraction of the 64 shared starts that reach the global minimum against the force calls spent, for every code and arm of Table~\ref{tbl:frozen-lj} and of the ablation in the Supporting Information, LJ\(_{38}\) on the left and LJ\(_{75}\) on the right.}
\label{fig:anytime}
\end{figure}

Table \ref{tbl:frozen-lj} runs anneal, GMIN, ABCluster,
and plain basin hopping in the anneal binary from the same 64
OptBench starts, under a protocol fixed before any search. On
LJ\(_{38}\) full anneal, GMIN, and ABCluster each hit 64/64, and
plain basin hopping hits 62/64. The median PAR-2 is 6591 force
calls for anneal, 14505 for GMIN, and 2571137 for ABCluster.
The paired test against GMIN gives a Holm-adjusted \(p\) of 0.091,
so anneal matches the hit rate of GMIN at a lower median cost the
test does not separate. On LJ\(_{75}\) anneal and GMIN both hit
64/64, at median PAR-2 of 30850.5 and 41210 force calls, and the
paired test gives a Holm-adjusted \(p\) of 1. Plain basin hopping
hits 1/64, a Holm-adjusted \(p\) of \(6.5\times10^{-18}\) on the
hits. ABCluster hits none of the 6 seeds that finished, each after
about \(4.3\times10^{7}\) calls at \(-396.28224883\), the icosahedral
minimum, and the other 58 seeds were still running. Figure
\ref{fig:anytime} plots the fraction of seeds that hit
against the force calls spent.

The Supporting Information switches off one component of full
anneal at a time on the same starts. Only the orbit move
separates. Without it the median PAR-2 of LJ\(_{75}\) rises from
30850.5 to 79341 force calls at 63/64 hits, a median paired rise of
36319 calls, Holm-adjusted \(p=1.8\times10^{-5}\). On LJ\(_{38}\)
the orbit move costs calls instead, a median paired 698.5,
Holm-adjusted \(p=2.4\times10^{-4}\). With Thompson allocation, the
return screen, or the predictive screen switched off, both systems
still hit 64/64, and no paired difference survives the Holm
correction. Independent chains match full anneal whenever the
first chain hits inside its quarter of the ceiling, so that arm
does not measure communication.

\begin{table}[t]
\centering
\scriptsize
\setlength{\tabcolsep}{3pt}
\caption{LJ\(_{38}\) and LJ\(_{75}\) under a protocol frozen before any search, on anneal develop f29dedc0. Every code reads the same 64 OptBench starts, and a run hits below the reference plus 0.001 within a ceiling of 1200000 force calls on LJ\(_{38}\) and 400000 on LJ\(_{75}\). GMIN runs the OptBench symmetrisation keyword files, and ABCluster 3.4 runs SN = 100, gmax = 1000 and glimit = 5. PAR-2 counts force calls, a miss costing twice the ceiling. The Holm-adjusted p values set each code against full anneal, by the exact McNemar test on the hits and the Wilcoxon signed-rank test on PAR-2, over every pair the campaign deposited. On LJ\(_{75}\) six ABCluster seeds had finished, and the other 58 were still running.}
\label{tbl:frozen-lj}
\begin{adjustbox}{max width=\textwidth}
\begin{tabular}{@{}llllrrll@{}}
\toprule
System & Code & Hits & 95\% interval & Median PAR-2 & Median minimisations & Holm \(p\), hits & Holm \(p\), PAR-2\\
\midrule
LJ\(_{38}\) & anneal, full & 64/64 & 0.943--1.000 & 6591 & 38.5 & -- & --\\
 & GMIN, symmetrisation & 64/64 & 0.943--1.000 & 14505 & 110.5 & 1 & 0.091\\
 & basin hopping, same binary & 62/64 & 0.893--0.991 & 310034.5 & 2502.5 & 1 & \(1.1\times10^{-10}\)\\
 & ABCluster 3.4 & 64/64 & 0.943--1.000 & 2571137 & not reported & 1 & \(1.1\times10^{-10}\)\\
\addlinespace
LJ\(_{75}\) & anneal, full & 64/64 & 0.943--1.000 & 30850.5 & 91 & -- & --\\
 & GMIN, symmetrisation & 64/64 & 0.943--1.000 & 41210 & 261.5 & 1 & 1\\
 & basin hopping, same binary & 1/64 & 0.003--0.083 & 800000 & 3560.5 & \(6.5\times10^{-18}\) & \(1.1\times10^{-10}\)\\
 & ABCluster 3.4 & 0/6 finished & 0.000--0.390 & -- & not reported & 0.81 & 0.53\\
\bottomrule
\end{tabular}
\end{adjustbox}
\end{table}

\subsection{Against ABCluster}
\label{sec:abcluster}
Table \ref{tbl:abcluster} sets anneal against ABCluster
\cite{zhangABClusterArtificialBee2015} on 6 more
systems, 32 paired seeds each, at the budget of each row. On
LJ\(_{55}\) both codes hit 32/32, and anneal spends fewer calls,
\(p=5.245\times10^{-6}\). On M\(_{55}\) at \(\rho=6\) anneal hits
32/32 and ABCluster 7/32. On M\(_{80}\) anneal hits 6/32 at
\(\rho=6\) and 10/32 at \(\rho=14\), and ABCluster hits none. On
LJ\(_{98}\) anneal hits 2/32 and ABCluster 1/32, \(p=0.285\).
Neither code hits M\(_{55}\) at \(\rho=14\) inside its budget.

\begin{table}[t]
\centering
\scriptsize
\setlength{\tabcolsep}{3pt}
\caption{anneal against ABCluster 3.4 on 32 paired seeds, 9200 to 9231, from random starts, one thread per run. anneal is \texttt{Config::for\_cluster} with the dual-core orbit completion on commit db22b99d. ABCluster is rebuilt from the 3.4 source with a seed, a call counter and a stop between local minimisations, at population 100, scout limit 5 and gmax 100000. The budget counts energy evaluations, and a miss costs twice the budget in PAR-2. A median first hit is not observed below 17 hits. The Wilcoxon test is two-sided on the 32 paired PAR-2 differences, ABCluster less anneal, and is undefined when both codes miss every seed. The LJ\(_{98}\) hit rule reads the Doye table, \(-543.642957\); anneal stores \(-543.665361\) for LJ\(_{98}\), and both of its hits end at \(-543.66536077\).}
\label{tbl:abcluster}
\begin{adjustbox}{max width=\textwidth}
\begin{tabular}{@{}lrllllrrl@{}}
\toprule
 & & \multicolumn{2}{c}{Hits} & \multicolumn{2}{c}{Median first hit (calls)} & \multicolumn{2}{c}{Mean PAR-2} & \\
\cmidrule(lr){3-4}\cmidrule(lr){5-6}\cmidrule(lr){7-8}
System & Budget & anneal & ABCluster & anneal & ABCluster & anneal & ABCluster & Wilcoxon \(p\)\\
\midrule
LJ\(_{55}\) & 800000 & 32/32 & 32/32 & 7410.5 & 41347.5 & 10420.3 & 75097.6 & \(5.245\times10^{-6}\)\\
LJ\(_{98}\) & 2000000 & 2/32 & 1/32 & not observed & not observed & 3762902.4 & 3897885.1 & 0.285\\
M\(_{55}\), \(\rho=6\) & 800000 & 32/32 & 7/32 & 7715.0 & not observed & 16223.1 & 1369731.4 & \(4.657\times10^{-10}\)\\
M\(_{55}\), \(\rho=14\) & 1500000 & 0/32 & 0/32 & not observed & not observed & 3000000.0 & 3000000.0 & undefined\\
M\(_{80}\), \(\rho=6\) & 1000000 & 6/32 & 0/32 & not observed & not observed & 1720869.6 & 2000000.0 & 0.02771\\
M\(_{80}\), \(\rho=14\) & 2000000 & 10/32 & 0/32 & not observed & not observed & 2890238.8 & 4000000.0 & 0.005062\\
\bottomrule
\end{tabular}
\end{adjustbox}
\end{table}
\subsection{Morse clusters against GMIN}
\label{sec:morse-gmin}
Table \ref{tbl:morse-rho} runs anneal and GMIN on the
Cambridge Morse table
\cite{doyeStructuralConsequencesRange1997} from \(N=20\)
to 80 at four ranges, 244 targets, with 16 paired seeds of each at
100000 calls. anneal reaches 145 targets and GMIN 131. anneal
reaches more at \(\rho=3, 10\), and 14, with 58, 22, and 19
against 54, 14, and 4. GMIN reaches more at \(\rho=6\), 59 against 46.
Over all targets GMIN has the lower mean PAR-2, 134409 calls
against 138251, and the more seed hits, 1439 against 1321. Neither
code reaches the primed minima 30D\('\), 61B\('\), 68D\('\), and
78D\('\). The Supporting Information lists every target.

\begin{table}[t]
\centering
\scriptsize
\setlength{\tabcolsep}{4pt}
\caption{anneal against GMIN on the Cambridge Morse table, \(N=20\) to 80 at each range \(\rho\), 16 paired seeds, 12000 to 12015, at a budget of 100000 energy-and-gradient calls per seed for each code. A target is one pair of \(N\) and \(\rho\), reached when one seed ends strictly below the Cambridge energy plus \(10^{-4}\). anneal is \texttt{Config::for\_cluster} with the dual-core orbit completion on commit db22b99d, and GMIN the Wales tarball with the Morse data file of its manual. The PAR-2 column is the mean over the targets of the mean PAR-2 of each, in calls. The Supporting Information gives every target.}
\label{tbl:morse-rho}
\begin{adjustbox}{max width=\textwidth}
\begin{tabular}{@{}lrrrrrr@{}}
\toprule
 & \multicolumn{2}{c}{Targets reached} & \multicolumn{2}{c}{Seed hits} & \multicolumn{2}{c}{Mean PAR-2} \\
\cmidrule(lr){2-3}\cmidrule(lr){4-5}\cmidrule(lr){6-7}
Range & anneal & GMIN & anneal & GMIN & anneal & GMIN\\
\midrule
\(\rho=3\) & 58/61 & 54/61 & 791/976 & 693/976 & 45156 & 65084\\
\(\rho=6\) & 46/61 & 59/61 & 378/976 & 704/976 & 133054 & 79025\\
\(\rho=10\) & 22/61 & 14/61 & 101/976 & 35/976 & 183088 & 194572\\
\(\rho=14\) & 19/61 & 4/61 & 51/976 & 7/976 & 191707 & 198954\\
\midrule
All four & 145/244 & 131/244 & 1321/3904 & 1439/3904 & 138251 & 134409\\
\bottomrule
\end{tabular}
\end{adjustbox}
\end{table}
\section{Conclusions}
\label{sec:conclusions}
From 64 shared starts anneal hits the global minimum of
LJ\(_{38}\) and LJ\(_{75}\) in 64 of 64 runs, as GMIN with
core-orbit symmetrisation does, at a lower median cost that the
paired test does not separate. The orbit move is the one component
that changes the cost on those two systems. On the Cambridge Morse
table anneal reaches 145 of 244 targets and GMIN 131, and GMIN
reaches more at \(\rho=6\). Against ABCluster anneal hits more seeds
of M\(_{55}\) at \(\rho=6\) and of M\(_{80}\).

The chains of anneal communicate. Each chain draws its next move
from a Normal-Gamma posterior on the depth that move reaches. It
closes the census of its packings by a Good-Turing occupancy law
\cite{goodPopulationFrequenciesSpecies1953}, and a Lean
derivation shows its hatch, its visit floor, and its retirement
rule. A
stalled chain symmetrises its cluster onto the point group it
nearly has. The chains share minima through an annealed-flow bank
under a held merge distance. With the per-chain ledger we compare
every code at an equal budget. An external engine reaches the
search through an rgpot bridge
\cite{goswamiOmniPotentRPCRgpot2026}.
\section{Author Contributions}
\label{sec:contributions}
Rohit Goswami: conceptualization, methodology, software, formal
analysis, Writing -- original draft and Writing -- review \& editing. Ruhila
Goswami: software, visualization, Writing -- review \& editing.
Amrita Goswami: methodology, Writing -- review \& editing. Moritz
Sallermann: software, methodology, Writing -- review \& editing.
Sonaly Goswami: validation, Writing -- review \& editing. Debabrata
Goswami: supervision, Writing -- review \& editing.
\section{Acknowledgments}
\label{sec:acks}
Rohit Goswami thanks his pets Arí, Yoda, Jude, Crystee, Thor,
and Loki, and the garden cats. In memoriam TuiTui, Snoopy, and
Cosmo. He thanks his cactus Ami and his in-laws Nagaselvamani and
Sankaranarayan. Ruhila Goswami set the figure colour palette.
\section{Declaration of generative AI and AI-assisted technologies in the writing process}
\label{sec:ai}
Chatbots were used to refine certain turns of phrase and to speed through some implementation details; all the work is the authors' own.
\section{Data and Software Availability}
\label{sec:availability}
anneal is under the MIT licence at
\url{https://github.com/HaoZeke/anneal}, with its documentation at
\url{https://anneal.rgoswami.me}. The LJ\(_{38}\) and LJ\(_{75}\)
campaign and the ablation ran its develop commit f29dedc0, the
campaigns against ABCluster and against GMIN on the Morse table its
main commit db22b99d with the orbit-completion patch their result
files carry. External engines reach anneal through rgpot
\cite{goswamiOmniPotentRPCRgpot2026}, under the MIT
licence at \url{https://github.com/OmniPotentRPC/rgpot}. The
repository \url{https://github.com/HaoZeke/simmAnnealTeX} holds the
result files of the three campaigns with the digest of each, the
script that renders every result table and Figure
\ref{fig:summary} from them, and a table that maps each
printed number to its file and field. The Supporting Information
gives the command of each campaign. Run from the folder that holds
the Snakefile, the first command below regenerates every result
table and the number map, and the second rebuilds this article, its
Supporting Information, and the preprint from their Org sources:

\begin{verbatim}
snakemake --cores 1 v1_tables
snakemake --cores 1 jcc --config org_load_path=/path/to/org-9.8.10
\end{verbatim}

\appendix
\section*{Supporting Information}
The following appendices are the journal Supporting Information, included here for the preprint.

\section{Algebra held out}
\label{sec:algebra}
The visiting width in the generalized step is

\begin{equation}
\begin{aligned}
\ln\sigma &= \frac{q_v-1}{3-q_v}\Big(
\tfrac12\ln\pi + (4-q_v)\ln(q_v-1)
- \frac{2-q_v}{q_v-1}\ln 2 - \ln(3-q_v)\\
&\qquad + \frac{\ln T}{q_v-1}
- \ln\Gamma\bigl(\tfrac{1}{q_v-1}-\tfrac12\bigr)
\Big),
\end{aligned}
\label{eq:log-sigma}
\end{equation}

with \(1<q_v<3\). A direct step whose absolute value is at most
\(10^{8}\) is kept. When that form is not finite, or its absolute
value exceeds \(10^{8}\), the step is rebuilt from the logarithm
of the same product. If that logarithm is not a number, or it
exceeds \(\ln 10^{8}\), the magnitude is drawn uniformly on
\((0,10^{8}]\) and carries the sign of the numerator normal.
Otherwise the magnitude is the exponential of the logarithm.

The temperature window in the main text is the branch on which
the lower edge lies below the upper edge. The gap used in the
window is at least \(10^{-12}\), and the remaining budget is at
least \(1\). The barrier starts at \(0\). It is an exponential
moving average of the uphill steps the chain rejected, with
decay \(0.98\). An accepted uphill step does not enter it. When
the lower edge is at least the upper edge, the temperature is
the larger of the lower edge and \(10^{-6}\), and the step is
counted as escape-forced.

The burst probabilities are in the main text. Under the cap at
eight relocations the expectation is

\begin{equation}
\mathbb{E}[K]
=\sum_{k=1}^{7} k\cdot\tfrac13\bigl(\tfrac23\bigr)^{k-1}
+8\bigl(\tfrac23\bigr)^{7}
=6305/2187.
\label{eq:burst-mean}
\end{equation}

The same law continued without the stop at eight has mean \(3\).
\section{Lean declarations}
\label{sec:lean}
Table \ref{tbl:lean} names the file that proves each
identity stated in the main text. Each file is under
\texttt{proofs/lean} at tag v0.11.1 of anneal.

\begin{table}[t]
\centering
\footnotesize
\setlength{\tabcolsep}{4pt}
\caption{Lean declarations for the identities in the main text, at tag v0.11.1 of anneal.}
\label{tbl:lean}
\begin{adjustbox}{max width=\textwidth}
\begin{tabular}{@{}>{\raggedright\arraybackslash}p{4.2cm}>{\raggedright\arraybackslash}p{12.2cm}@{}}
\toprule
File & Theorems\\
\midrule
\path{Hop/OccupancyGt.lean} & \path{leftover_hatch}, \path{hatch_next_one_fifth_sign}, \path{leftover_nick_then_rise}, \path{three_singletons_stable_at_twenty}, \path{three_singletons_unstable_at_nineteen}, \path{visits_floor_three_fifths}, \path{production_min_visits}, \path{cheeger_sq_gap}, \path{cheeger_cap}, \path{fes_factor_minus_one}, \path{fes_scale_invariant}, \path{fes_undefined_one_family}, \path{mixed_threshold_bw}, \path{certify_min_samples}, \path{mixing_alone_does_not_retire}, \path{one_book_family_waives_leftover}\\
\path{Hop/CostScreen.lean} & \path{threshold_identity}, \path{measured_hop_seven_fifteenths}\\
\path{Gpmd/CriticalTemp.lean} & \path{inv_theta_half_factor}, \path{factor_pos_at_half}, \path{factor_zero_at_two}, \path{factor_neg_at_four}, \path{theta_star_in_window}\\
\path{Hop/SeamLadder.lean} & \path{alone_step}, \path{cloned_const}, \path{cloning_dominates}, \path{road_priced}\\
\path{Hop/TempLadder.lean} & \path{ladder_pays}, \path{ladder_gain}\\
\path{Hop/SharedChannel.lean} & \path{sharper}, \path{fivefold_beats_positional}, \path{broadcast_scales}, \path{HistoryHook_never_coordinates}\\
\path{Gpmd/DensityRatio.lean} & pairing identities of the drift factor\\
\bottomrule
\end{tabular}
\end{adjustbox}
\end{table}
\section{Benchmark specification}
\label{sec:spec}
Table \ref{tbl:spec} gives, for each campaign, its systems, seeds, starts, budget per seed and hit
tolerance. The Lennard-Jones surface is the 12-6 form of Jones
\cite{jonesDeterminationMolecularFields1924}, and the
Morse surface that of Morse
\cite{morseDiatomicMoleculesAccording1929} at the ranges
of Doye and Wales
\cite{doyeStructuralConsequencesRange1997}. Every Morse
reference is the lowest minimum of the Cambridge Cluster Database
table at its size and range
\cite{walesCambridgeClusterDatabase}.

\begin{table}[t]
\centering
\scriptsize
\setlength{\tabcolsep}{2.5pt}
\caption{The three campaigns. Energies are in units of the pair well depth for the Lennard-Jones and Morse surfaces. A run hits strictly below the reference plus the tolerance.}
\label{tbl:spec}
\begin{adjustbox}{max width=\textwidth}
\begin{tabular}{@{}>{\raggedright\arraybackslash}p{2.4cm}>{\raggedright\arraybackslash}p{3.2cm}>{\raggedright\arraybackslash}p{1.6cm}>{\raggedright\arraybackslash}p{3.0cm}>{\raggedright\arraybackslash}p{2.4cm}>{\raggedright\arraybackslash}p{1.8cm}@{}}
\toprule
Campaign & Systems & Seeds & Start & Budget per seed (force calls) & Hit tolerance\\
\midrule
Shared starts & LJ\(_{38}\), LJ\(_{75}\) & 0 to 63 & the OptBench start of each seed, shared by every code & 1200000 on LJ\(_{38}\), 400000 on LJ\(_{75}\) & 0.001\\
Against ABCluster & LJ\(_{55}\), LJ\(_{98}\); M\(_{55}\) and M\(_{80}\) at \(\rho=6\), 14 & 9200 to 9231 & random, by each code's own generator & the budget of each row & 0.001\\
Morse against GMIN & M\(_N\), \(N=20\) to 80, \(\rho=3\), 6, 10, 14 & 12000 to 12015 & random & 100000 & \(10^{-4}\)\\
\bottomrule
\end{tabular}
\end{adjustbox}
\end{table}
\section{Codes, versions, and builds}
\label{sec:codes}
Table \ref{tbl:codes} names, for each campaign, the
version or commit of each code, its build, the route of its
potential, and the counter from which its force calls are read.

\begin{table}[t]
\centering
\scriptsize
\setlength{\tabcolsep}{2.5pt}
\caption{The codes of each campaign, their pins and builds, the route of the potential, and the counter of force calls. The orbit-completion patch of anneal and the counting patches of ABCluster and GMIN are in the result files of their campaigns; neither counting patch changes a proposal, an acceptance step or a minimiser setting.}
\label{tbl:codes}
\begin{adjustbox}{max width=\textwidth}
\begin{tabular}{@{}>{\raggedright\arraybackslash}p{2.0cm}>{\raggedright\arraybackslash}p{1.5cm}>{\raggedright\arraybackslash}p{3.6cm}>{\raggedright\arraybackslash}p{3.4cm}>{\raggedright\arraybackslash}p{2.0cm}>{\raggedright\arraybackslash}p{2.4cm}@{}}
\toprule
Campaign & Code & Pin & Build & Potential & Force calls read from\\
\midrule
Shared starts & anneal & develop f29dedc0 & rustc 1.88.0, release, default features & built in & per-chain ledger\\
 & GMIN & OptBench wales-27058 & gfortran 13.3.0 & its own & the function calls of its hit line\\
 & ABCluster & 3.4 Linux binary of 6 January 2026 & as distributed & its own & a counter on its Lennard-Jones gradient\\
\addlinespace
Against ABCluster & anneal & main db22b99d and the orbit-completion patch & rustc 1.91.0, release & built in & per-chain ledger\\
 & ABCluster & 3.4 source & g++ 13.3.0 at the distribution flags, with a seed, a call counter and a stop between minimisations & its own & that counter\\
\addlinespace
Morse against GMIN & anneal & main db22b99d and the orbit-completion patch & rustc 1.99.1, release & built in & per-chain ledger\\
 & GMIN & Wales source tarball of 31 October 2026 & gfortran 13.3.0, CMake 3.28.3, Release, with a print of its call count on each quench & its own & that print\\
\bottomrule
\end{tabular}
\end{adjustbox}
\end{table}
\section{Reproduction}
\label{sec:reproduction}
Each campaign fixed its protocol before its first seed, and each
per-seed record carries the command that ran it. The LJ\(_{38}\)
and LJ\(_{75}\) campaign and the ablation build anneal at develop
commit f29dedc0 with

\begin{verbatim}
cargo build --release --locked --example ablation_campaign
ablation_campaign <system> <arm> <seed> <start.npy> <record.json>
\end{verbatim}

where the start is the shared OptBench start of that seed and the
arm is \texttt{full}, \texttt{plain\_bh} or one switch off. GMIN runs as \texttt{GMIN} in
a folder that holds the OptBench data file and that start, and
ABCluster 3.4 as \texttt{atom job.inp} at SN = 100, gmax = 1000 and
glimit = 5. The campaign against ABCluster builds anneal at main
commit db22b99d with the orbit-completion patch of its result files
and runs

\begin{verbatim}
cargo build --release --example h2h_arm3
h2h_arm3 <system> <budget> <seed>
\end{verbatim}

and the Morse campaign against GMIN runs, on the same commit and
patch,

\begin{verbatim}
cargo build --release --example morse_h2h
morse_h2h <rho> <n> 100000 <seed> <reference>
\end{verbatim}

for seeds 12000 to 12015, with GMIN built from the Wales source
tarball and run as \texttt{GMIN} on the Morse data file of its manual. Run from the
folder of the manuscript repository that holds the Snakefile, the
first command below regenerates every result table of both
documents, the summary figure, and the map from each printed
number to its file and field, and the second rebuilds the main
text, this document, and the preprint from their Org sources:

\begin{verbatim}
snakemake --cores 1 v1_tables
snakemake --cores 1 jcc --config org_load_path=/path/to/org-9.8.10
\end{verbatim}

The rule \texttt{jcc\_tex} exports each Org file with Org 9.8.10,
\texttt{jcc\_pdf} compiles it with tectonic against the Wiley NJD class
that the tree carries, and \texttt{jcc\_arxiv} stages and compiles the
preprint.
\section{Ablation of the components}
\label{sec:ablation}
Table \ref{tbl:ablation} switches off one component of
full anneal at a time, on the 64 shared OptBench starts of LJ\(_{38}\)
and LJ\(_{75}\) and under the frozen protocol. The
arm \texttt{no\_thompson} turns off Thompson allocation and the
depth reward, \texttt{no\_rscreen} the return screen,
\texttt{no\_predictive} the posterior-predictive screen, and
\texttt{no\_orbit} the orbit move. The arm \texttt{independent}
runs four chains with no shared state at a quarter of the ceiling
each, and a seed hits when any chain hits. Only the orbit move
separates from full anneal. It saves a median of 36319 paired calls
on LJ\(_{75}\) and costs 698.5 on LJ\(_{38}\). Thompson allocation,
the return screen, the predictive screen, and the independent arm
show no difference after the Holm correction. Every arm draws from
the all, single, surface, shell and sym moves of the communicating
preset. In \texttt{no\_rscreen} the recognition stand-in of that
preset stays on, so a trial that falls back into a basin already
recorded still skips its quench, and the arm measures only what the
return screen adds beyond recognition. The other systems of the
protocol are outside this ablation.

\begin{table}[t]
\centering
\scriptsize
\setlength{\tabcolsep}{3pt}
\caption{Full anneal with one switch off, on the 64 OptBench starts of the LJ\(_{38}\) and LJ\(_{75}\) table, anneal develop f29dedc0, at ceilings of 1200000 force calls on LJ\(_{38}\) and 400000 on LJ\(_{75}\). Median PAR-2 counts force calls, a miss costing twice the ceiling. The paired difference is the median over the 64 seeds of the PAR-2 of the arm less that of full anneal, so a positive difference means the switch saves calls. The Holm correction runs over the ten Wilcoxon signed-rank tests of this table. No hit count differs from full anneal after Holm.}
\label{tbl:ablation}
\begin{adjustbox}{max width=\textwidth}
\begin{tabular}{@{}lrrrlrrrl@{}}
\toprule
 & \multicolumn{4}{c}{LJ\(_{38}\)} & \multicolumn{4}{c}{LJ\(_{75}\)} \\
\cmidrule(lr){2-5}\cmidrule(lr){6-9}
Arm & Hits & PAR-2 & Paired difference & Holm \(p\) & Hits & PAR-2 & Paired difference & Holm \(p\)\\
\midrule
\texttt{full} & 64/64 & 6591 & -- & -- & 64/64 & 30850.5 & -- & --\\
\texttt{no\_thompson} & 64/64 & 6634.5 & 43.5 & 0.89 & 64/64 & 32893 & 312 & 1\\
\texttt{no\_rscreen} & 64/64 & 3841.5 & 34 & 0.81 & 64/64 & 32158.5 & \(-256.5\) & 1\\
\texttt{no\_predictive} & 64/64 & 6334.5 & \(-65\) & 0.11 & 64/64 & 26265.5 & \(-181.5\) & 1\\
\texttt{no\_orbit} & 64/64 & 7363 & \(-698.5\) & \(2.4\times10^{-4}\) & 63/64 & 79341 & 36319 & \(1.8\times10^{-5}\)\\
\texttt{independent} & 64/64 & 6591 & 0 & 1 & 64/64 & 30850.5 & 0 & 1\\
\bottomrule
\end{tabular}
\end{adjustbox}
\end{table}
\section{Morse clusters against GMIN, target by target}
\label{sec:morse-seeds}
Table \ref{tbl:morse-n} gives every target of the
comparison of anneal with GMIN on the Cambridge Morse table
\cite{doyeStructuralConsequencesRange1997}, one row for
each cluster size \(N\) and range \(\rho\). GMIN hits 23 targets
on which anneal hits none of its 16 seeds, 13 of them at
\(\rho=6\).

\begingroup
\scriptsize
\setlength{\tabcolsep}{4pt}
\begin{longtable}{@{}rrlrrrrll@{}}
\caption{anneal against GMIN on every target of the Cambridge Morse table, 16 paired seeds, 12000 to 12015, 100000 energy-and-gradient calls per seed, as in the Morse table. The minimum is the Cambridge label of the target, and a prime marks the four that neither code reaches. Hits are of the 16 seeds and PAR-2 is the mean in calls. The exact McNemar test sets the paired hits against each other and the Wilcoxon signed-rank test the paired PAR-2, both two-sided, and a dash marks a Wilcoxon test with no nonzero difference.}\label{tbl:morse-n}\\
\toprule
 & & & \multicolumn{2}{c}{Hits} & \multicolumn{2}{c}{Mean PAR-2} & & \\
\(\rho\) & \(N\) & Minimum & anneal & GMIN & anneal & GMIN & McNemar \(p\) & Wilcoxon \(p\)\\
\midrule
\endfirsthead
\toprule
\(\rho\) & \(N\) & Minimum & anneal & GMIN & anneal & GMIN & McNemar \(p\) & Wilcoxon \(p\)\\
\midrule
\endhead
\bottomrule
\endfoot
3 & 20 & 20A & 16/16 & 16/16 & 164 & 162 & 1 & 1\\
3 & 21 & 21A & 16/16 & 16/16 & 892 & 372 & 1 & 0.025\\
3 & 22 & 22A & 16/16 & 16/16 & 232 & 198 & 1 & 0.71\\
3 & 23 & 23A & 16/16 & 16/16 & 384 & 152 & 1 & 0.051\\
3 & 24 & 24A & 16/16 & 16/16 & 407 & 246 & 1 & 0.14\\
3 & 25 & 25A & 16/16 & 16/16 & 255 & 201 & 1 & 0.32\\
3 & 26 & 26A & 16/16 & 16/16 & 264 & 213 & 1 & 0.77\\
3 & 27 & 27A & 16/16 & 16/16 & 403 & 198 & 1 & 0.12\\
3 & 28 & 28A & 16/16 & 16/16 & 651 & 317 & 1 & 0.38\\
3 & 29 & 29A & 16/16 & 16/16 & 568 & 267 & 1 & 0.083\\
3 & 30 & 30A & 16/16 & 16/16 & 4325 & 1226 & 1 & \(1.7\times10^{-3}\)\\
3 & 31 & 31A & 16/16 & 16/16 & 3265 & 1737 & 1 & 0.27\\
3 & 32 & 32A & 16/16 & 16/16 & 6095 & 914 & 1 & \(4.3\times10^{-4}\)\\
3 & 33 & 33A & 16/16 & 16/16 & 7276 & 1129 & 1 & \(1.5\times10^{-4}\)\\
3 & 34 & 34A & 16/16 & 16/16 & 1036 & 346 & 1 & \(7.6\times10^{-3}\)\\
3 & 35 & 35A & 16/16 & 16/16 & 3491 & 1221 & 1 & 0.039\\
3 & 36 & 36A & 16/16 & 16/16 & 9265 & 24037 & 1 & 0.065\\
3 & 37 & 37A & 16/16 & 16/16 & 6816 & 2762 & 1 & 0.018\\
3 & 38 & 38A & 16/16 & 16/16 & 2106 & 1579 & 1 & 0.98\\
3 & 39 & 39A & 16/16 & 16/16 & 8743 & 2559 & 1 & \(2.7\times10^{-3}\)\\
3 & 40 & 40A & 16/16 & 16/16 & 4620 & 1044 & 1 & \(4.3\times10^{-4}\)\\
3 & 41 & 41A & 16/16 & 16/16 & 3048 & 589 & 1 & \(1.3\times10^{-3}\)\\
3 & 42 & 42A & 16/16 & 16/16 & 2601 & 610 & 1 & 0.083\\
3 & 43 & 43A & 16/16 & 16/16 & 1846 & 821 & 1 & 0.018\\
3 & 44 & 44A & 16/16 & 16/16 & 2172 & 497 & 1 & \(1.0\times10^{-3}\)\\
3 & 45 & 45A & 16/16 & 16/16 & 1245 & 846 & 1 & 0.32\\
3 & 46 & 46A & 16/16 & 8/16 & 2157 & 124330 & \(7.8\times10^{-3}\) & \(3.1\times10^{-5}\)\\
3 & 47 & 47A & 16/16 & 12/16 & 2305 & 77028 & 0.12 & \(6.1\times10^{-5}\)\\
3 & 48 & 48A & 16/16 & 12/16 & 1040 & 68663 & 0.12 & \(3.1\times10^{-5}\)\\
3 & 49 & 49A & 16/16 & 15/16 & 2165 & 26855 & 1 & \(3.1\times10^{-5}\)\\
3 & 50 & 50A & 16/16 & 16/16 & 8388 & 9517 & 1 & 0.63\\
3 & 51 & 51A & 16/16 & 16/16 & 15107 & 26204 & 1 & 0.3\\
3 & 52 & 52A & 16/16 & 16/16 & 1640 & 1946 & 1 & 0.35\\
3 & 53 & 53A & 16/16 & 16/16 & 3401 & 1154 & 1 & 0.14\\
3 & 54 & 54A & 16/16 & 16/16 & 4122 & 12658 & 1 & \(1.5\times10^{-4}\)\\
3 & 55 & 55A & 16/16 & 2/16 & 6331 & 182001 & \(1.2\times10^{-4}\) & \(3.1\times10^{-5}\)\\
3 & 56 & 56A & 16/16 & 15/16 & 7529 & 51139 & 1 & \(3.1\times10^{-4}\)\\
3 & 57 & 57A & 16/16 & 16/16 & 4790 & 5638 & 1 & 0.74\\
3 & 58 & 58A & 16/16 & 16/16 & 5534 & 2843 & 1 & 0.32\\
3 & 59 & 59A & 13/16 & 14/16 & 75555 & 41742 & 1 & 0.022\\
3 & 60 & 60A & 16/16 & 16/16 & 7918 & 6297 & 1 & 0.9\\
3 & 61 & 61A & 16/16 & 16/16 & 12818 & 2984 & 1 & \(1.7\times10^{-3}\)\\
3 & 62 & 62A & 16/16 & 6/16 & 25457 & 153299 & \(2.0\times10^{-3}\) & \(3.1\times10^{-5}\)\\
3 & 63 & 63A & 16/16 & 3/16 & 26912 & 172198 & \(2.4\times10^{-4}\) & \(6.1\times10^{-5}\)\\
3 & 64 & 64A & 15/16 & 4/16 & 36496 & 163698 & \(9.8\times10^{-4}\) & \(1.8\times10^{-4}\)\\
3 & 65 & 65A & 16/16 & 11/16 & 19588 & 88470 & 0.062 & \(5.2\times10^{-3}\)\\
3 & 66 & 66A & 16/16 & 11/16 & 16936 & 91655 & 0.062 & \(4.3\times10^{-4}\)\\
3 & 67 & 67A & 10/16 & 1/16 & 98065 & 193471 & 0.012 & \(2.0\times10^{-3}\)\\
3 & 68 & 68A & 2/16 & 3/16 & 178449 & 173702 & 1 & 1\\
3 & 69 & 69A & 3/16 & 1/16 & 176226 & 191945 & 0.62 & 0.62\\
3 & 70 & 70A & 1/16 & 2/16 & 192394 & 180793 & 1 & 0.5\\
3 & 71 & 71A & 5/16 & 3/16 & 153788 & 165899 & 0.69 & 0.94\\
3 & 72 & 72A & 10/16 & 9/16 & 111554 & 117223 & 1 & 0.79\\
3 & 73 & 73A & 2/16 & 1/16 & 180918 & 192540 & 1 & 0.5\\
3 & 74 & 74A & 0/16 & 0/16 & 200000 & 200000 & 1 & --\\
3 & 75 & 75A & 2/16 & 0/16 & 180291 & 200000 & 0.5 & 0.5\\
3 & 76 & 76A & 5/16 & 0/16 & 148732 & 200000 & 0.062 & 0.062\\
3 & 77 & 77A & 1/16 & 0/16 & 190595 & 200000 & 1 & 1\\
3 & 78 & 78A & 0/16 & 0/16 & 200000 & 200000 & 1 & --\\
3 & 79 & 79A & 2/16 & 0/16 & 185120 & 200000 & 0.5 & 0.5\\
3 & 80 & 80A & 0/16 & 0/16 & 200000 & 200000 & 1 & --\\
6 & 20 & 20A & 16/16 & 16/16 & 4216 & 2402 & 1 & 0.12\\
6 & 21 & 21B & 16/16 & 16/16 & 9961 & 2319 & 1 & \(3.4\times10^{-3}\)\\
6 & 22 & 22A & 16/16 & 16/16 & 4875 & 3102 & 1 & 0.3\\
6 & 23 & 23A & 16/16 & 16/16 & 16975 & 7548 & 1 & 0.044\\
6 & 24 & 24A & 16/16 & 16/16 & 3890 & 3798 & 1 & 0.94\\
6 & 25 & 25A & 16/16 & 16/16 & 3859 & 3533 & 1 & 0.9\\
6 & 26 & 26A & 16/16 & 16/16 & 7973 & 10974 & 1 & 0.3\\
6 & 27 & 27B & 10/16 & 15/16 & 101510 & 33774 & 0.062 & \(1.2\times10^{-3}\)\\
6 & 28 & 28B & 16/16 & 15/16 & 24937 & 32201 & 1 & 0.9\\
6 & 29 & 29A & 16/16 & 15/16 & 17422 & 32009 & 1 & 0.71\\
6 & 30 & 30B & 12/16 & 13/16 & 84166 & 78719 & 1 & 0.94\\
6 & 31 & 31B & 16/16 & 8/16 & 32907 & 117269 & \(7.8\times10^{-3}\) & \(2.1\times10^{-3}\)\\
6 & 32 & 32B & 3/16 & 15/16 & 171957 & 45679 & \(4.9\times10^{-4}\) & \(1.2\times10^{-4}\)\\
6 & 33 & 33C & 2/16 & 15/16 & 182030 & 38831 & \(2.4\times10^{-4}\) & \(6.1\times10^{-5}\)\\
6 & 34 & 34B & 1/16 & 3/16 & 192521 & 174900 & 0.62 & 0.38\\
6 & 35 & 35C & 0/16 & 15/16 & 200000 & 48689 & \(6.1\times10^{-5}\) & \(6.1\times10^{-5}\)\\
6 & 36 & 36C & 4/16 & 14/16 & 165379 & 58861 & 0.013 & \(1.3\times10^{-3}\)\\
6 & 37 & 37B & 0/16 & 13/16 & 200000 & 70532 & \(2.4\times10^{-4}\) & \(2.4\times10^{-4}\)\\
6 & 38 & 38E & 7/16 & 10/16 & 144505 & 106463 & 0.55 & 0.17\\
6 & 39 & 39B & 10/16 & 16/16 & 96154 & 27824 & 0.031 & 0.034\\
6 & 40 & 40B & 11/16 & 15/16 & 102160 & 39746 & 0.12 & \(8.5\times10^{-4}\)\\
6 & 41 & 41B & 5/16 & 14/16 & 152752 & 60648 & \(3.9\times10^{-3}\) & \(5.2\times10^{-3}\)\\
6 & 42 & 42B & 1/16 & 15/16 & 191506 & 48850 & \(5.2\times10^{-4}\) & \(2.1\times10^{-4}\)\\
6 & 43 & 43B & 0/16 & 15/16 & 200000 & 40200 & \(6.1\times10^{-5}\) & \(6.1\times10^{-5}\)\\
6 & 44 & 44B & 0/16 & 16/16 & 200000 & 38899 & \(3.1\times10^{-5}\) & \(3.1\times10^{-5}\)\\
6 & 45 & 45B & 4/16 & 16/16 & 163073 & 41048 & \(4.9\times10^{-4}\) & \(3.1\times10^{-4}\)\\
6 & 46 & 46B & 3/16 & 15/16 & 171186 & 42946 & \(4.9\times10^{-4}\) & \(1.8\times10^{-4}\)\\
6 & 47 & 47B & 1/16 & 15/16 & 189180 & 46700 & \(1.2\times10^{-4}\) & \(1.2\times10^{-4}\)\\
6 & 48 & 48B & 2/16 & 16/16 & 180137 & 41610 & \(1.2\times10^{-4}\) & \(3.1\times10^{-5}\)\\
6 & 49 & 49B & 2/16 & 16/16 & 180562 & 32402 & \(1.2\times10^{-4}\) & \(3.1\times10^{-5}\)\\
6 & 50 & 50B & 4/16 & 14/16 & 164826 & 48409 & \(2.0\times10^{-3}\) & \(3.7\times10^{-4}\)\\
6 & 51 & 51B & 16/16 & 12/16 & 20746 & 95498 & 0.12 & \(9.2\times10^{-5}\)\\
6 & 52 & 52B & 16/16 & 15/16 & 22044 & 41070 & 1 & 0.3\\
6 & 53 & 53B & 16/16 & 16/16 & 12901 & 16736 & 1 & 0.35\\
6 & 54 & 54B & 16/16 & 16/16 & 17919 & 27312 & 1 & 0.43\\
6 & 55 & 55B & 16/16 & 15/16 & 14589 & 35447 & 1 & 0.083\\
6 & 56 & 56B & 16/16 & 14/16 & 24087 & 49788 & 0.5 & 0.1\\
6 & 57 & 57B & 12/16 & 16/16 & 96451 & 31108 & 0.12 & \(1.0\times10^{-3}\)\\
6 & 58 & 58B & 2/16 & 13/16 & 183177 & 72230 & \(9.8\times10^{-4}\) & \(4.9\times10^{-4}\)\\
6 & 59 & 59D & 1/16 & 13/16 & 193328 & 71024 & \(4.9\times10^{-4}\) & \(2.4\times10^{-4}\)\\
6 & 60 & 60B & 0/16 & 15/16 & 200000 & 49672 & \(6.1\times10^{-5}\) & \(6.1\times10^{-5}\)\\
6 & 61 & 61B & 0/16 & 13/16 & 200000 & 75672 & \(2.4\times10^{-4}\) & \(2.4\times10^{-4}\)\\
6 & 62 & 62B & 1/16 & 10/16 & 188665 & 92382 & \(3.9\times10^{-3}\) & \(3.9\times10^{-3}\)\\
6 & 63 & 63B & 2/16 & 14/16 & 181680 & 62564 & \(4.9\times10^{-4}\) & \(1.2\times10^{-4}\)\\
6 & 64 & 64B & 3/16 & 11/16 & 170340 & 97161 & 0.021 & 0.012\\
6 & 65 & 65C & 0/16 & 1/16 & 200000 & 190253 & 1 & 1\\
6 & 66 & 66C & 0/16 & 7/16 & 200000 & 134306 & 0.016 & 0.016\\
6 & 67 & 67C & 3/16 & 6/16 & 171789 & 135303 & 0.51 & 0.25\\
6 & 68 & 68B & 0/16 & 6/16 & 200000 & 145800 & 0.031 & 0.031\\
6 & 69 & 69D & 0/16 & 5/16 & 200000 & 160510 & 0.062 & 0.062\\
6 & 70 & 70D & 4/16 & 2/16 & 168194 & 184590 & 0.62 & 0.19\\
6 & 71 & 71C & 1/16 & 10/16 & 190328 & 107974 & \(3.9\times10^{-3}\) & \(3.9\times10^{-3}\)\\
6 & 72 & 72C & 1/16 & 5/16 & 193374 & 157979 & 0.22 & 0.062\\
6 & 73 & 73B & 2/16 & 7/16 & 180362 & 141862 & 0.12 & 0.2\\
6 & 74 & 74B & 0/16 & 8/16 & 200000 & 129200 & \(7.8\times10^{-3}\) & \(7.8\times10^{-3}\)\\
6 & 75 & 75C & 4/16 & 1/16 & 162099 & 191015 & 0.25 & 0.12\\
6 & 76 & 76D & 0/16 & 1/16 & 200000 & 192897 & 1 & 1\\
6 & 77 & 77C & 0/16 & 0/16 & 200000 & 200000 & 1 & --\\
6 & 78 & 78D\('\) & 0/16 & 0/16 & 200000 & 200000 & 1 & --\\
6 & 79 & 79B & 4/16 & 1/16 & 163630 & 192732 & 0.25 & 0.12\\
6 & 80 & 80B & 0/16 & 5/16 & 200000 & 157546 & 0.062 & 0.062\\
10 & 20 & 20A & 16/16 & 12/16 & 23325 & 73486 & 0.12 & 0.029\\
10 & 21 & 21C & 0/16 & 1/16 & 200000 & 188925 & 1 & 1\\
10 & 22 & 22B & 0/16 & 0/16 & 200000 & 200000 & 1 & --\\
10 & 23 & 23C & 0/16 & 1/16 & 200000 & 191224 & 1 & 1\\
10 & 24 & 24B & 0/16 & 2/16 & 200000 & 185269 & 0.5 & 0.5\\
10 & 25 & 25B & 0/16 & 6/16 & 200000 & 142122 & 0.031 & 0.031\\
10 & 26 & 26B & 0/16 & 0/16 & 200000 & 200000 & 1 & --\\
10 & 27 & 27D & 0/16 & 0/16 & 200000 & 200000 & 1 & --\\
10 & 28 & 28D & 0/16 & 4/16 & 200000 & 167047 & 0.12 & 0.12\\
10 & 29 & 29C & 0/16 & 0/16 & 200000 & 200000 & 1 & --\\
10 & 30 & 30C & 0/16 & 2/16 & 200000 & 185461 & 0.5 & 0.5\\
10 & 31 & 31D & 1/16 & 0/16 & 191058 & 200000 & 1 & 1\\
10 & 32 & 32B & 1/16 & 1/16 & 187799 & 190605 & 1 & 1\\
10 & 33 & 33E & 0/16 & 0/16 & 200000 & 200000 & 1 & --\\
10 & 34 & 34D & 0/16 & 0/16 & 200000 & 200000 & 1 & --\\
10 & 35 & 35D & 1/16 & 1/16 & 192070 & 190705 & 1 & 1\\
10 & 36 & 36E & 0/16 & 0/16 & 200000 & 200000 & 1 & --\\
10 & 37 & 37C & 5/16 & 1/16 & 157454 & 191633 & 0.22 & 0.22\\
10 & 38 & 38D & 11/16 & 0/16 & 91630 & 200000 & \(9.8\times10^{-4}\) & \(9.8\times10^{-4}\)\\
10 & 39 & 39C & 7/16 & 0/16 & 134145 & 200000 & 0.016 & 0.016\\
10 & 40 & 40C & 4/16 & 0/16 & 159969 & 200000 & 0.12 & 0.12\\
10 & 41 & 41D & 0/16 & 0/16 & 200000 & 200000 & 1 & --\\
10 & 42 & 42C & 0/16 & 0/16 & 200000 & 200000 & 1 & --\\
10 & 43 & 43D & 0/16 & 0/16 & 200000 & 200000 & 1 & --\\
10 & 44 & 44D & 0/16 & 0/16 & 200000 & 200000 & 1 & --\\
10 & 45 & 45D & 0/16 & 0/16 & 200000 & 200000 & 1 & --\\
10 & 46 & 46B & 0/16 & 0/16 & 200000 & 200000 & 1 & --\\
10 & 47 & 47C & 0/16 & 0/16 & 200000 & 200000 & 1 & --\\
10 & 48 & 48C & 0/16 & 0/16 & 200000 & 200000 & 1 & --\\
10 & 49 & 49C & 0/16 & 0/16 & 200000 & 200000 & 1 & --\\
10 & 50 & 50C & 0/16 & 0/16 & 200000 & 200000 & 1 & --\\
10 & 51 & 51D & 0/16 & 0/16 & 200000 & 200000 & 1 & --\\
10 & 52 & 52D & 0/16 & 0/16 & 200000 & 200000 & 1 & --\\
10 & 53 & 53B & 6/16 & 1/16 & 138335 & 187882 & 0.12 & 0.22\\
10 & 54 & 54B & 8/16 & 1/16 & 123130 & 191602 & 0.016 & 0.016\\
10 & 55 & 55B & 11/16 & 1/16 & 89836 & 191949 & \(2.0\times10^{-3}\) & \(9.8\times10^{-4}\)\\
10 & 56 & 56B & 9/16 & 1/16 & 95264 & 190983 & 0.021 & \(3.9\times10^{-3}\)\\
10 & 57 & 57C & 0/16 & 0/16 & 200000 & 200000 & 1 & --\\
10 & 58 & 58C & 1/16 & 0/16 & 187892 & 200000 & 1 & 1\\
10 & 59 & 59E & 0/16 & 0/16 & 200000 & 200000 & 1 & --\\
10 & 60 & 60D & 0/16 & 0/16 & 200000 & 200000 & 1 & --\\
10 & 61 & 61B\('\) & 0/16 & 0/16 & 200000 & 200000 & 1 & --\\
10 & 62 & 62F & 0/16 & 0/16 & 200000 & 200000 & 1 & --\\
10 & 63 & 63C & 0/16 & 0/16 & 200000 & 200000 & 1 & --\\
10 & 64 & 64C & 0/16 & 0/16 & 200000 & 200000 & 1 & --\\
10 & 65 & 65E & 1/16 & 0/16 & 193687 & 200000 & 1 & 1\\
10 & 66 & 66E & 0/16 & 0/16 & 200000 & 200000 & 1 & --\\
10 & 67 & 67D & 2/16 & 0/16 & 184263 & 200000 & 0.5 & 0.5\\
10 & 68 & 68D\('\) & 0/16 & 0/16 & 200000 & 200000 & 1 & --\\
10 & 69 & 69F & 0/16 & 0/16 & 200000 & 200000 & 1 & --\\
10 & 70 & 70E & 2/16 & 0/16 & 180352 & 200000 & 0.5 & 0.5\\
10 & 71 & 71D & 3/16 & 0/16 & 169136 & 200000 & 0.25 & 0.25\\
10 & 72 & 72D & 0/16 & 0/16 & 200000 & 200000 & 1 & --\\
10 & 73 & 73D & 0/16 & 0/16 & 200000 & 200000 & 1 & --\\
10 & 74 & 74C & 2/16 & 0/16 & 176451 & 200000 & 0.5 & 0.5\\
10 & 75 & 75C & 2/16 & 0/16 & 176628 & 200000 & 0.5 & 0.5\\
10 & 76 & 76E & 0/16 & 0/16 & 200000 & 200000 & 1 & --\\
10 & 77 & 77D & 5/16 & 0/16 & 144565 & 200000 & 0.062 & 0.062\\
10 & 78 & 78E & 1/16 & 0/16 & 190048 & 200000 & 1 & 1\\
10 & 79 & 79D & 2/16 & 0/16 & 181337 & 200000 & 0.5 & 0.5\\
10 & 80 & 80D & 0/16 & 0/16 & 200000 & 200000 & 1 & --\\
14 & 20 & 20B & 15/16 & 0/16 & 47942 & 200000 & \(6.1\times10^{-5}\) & \(6.1\times10^{-5}\)\\
14 & 21 & 21C & 0/16 & 2/16 & 200000 & 182628 & 0.5 & 0.5\\
14 & 22 & 22B & 0/16 & 3/16 & 200000 & 173835 & 0.25 & 0.25\\
14 & 23 & 23C & 0/16 & 1/16 & 200000 & 188727 & 1 & 1\\
14 & 24 & 24B & 0/16 & 0/16 & 200000 & 200000 & 1 & --\\
14 & 25 & 25B & 0/16 & 1/16 & 200000 & 191014 & 1 & 1\\
14 & 26 & 26B & 0/16 & 0/16 & 200000 & 200000 & 1 & --\\
14 & 27 & 27D & 0/16 & 0/16 & 200000 & 200000 & 1 & --\\
14 & 28 & 28D & 0/16 & 0/16 & 200000 & 200000 & 1 & --\\
14 & 29 & 29E & 2/16 & 0/16 & 178340 & 200000 & 0.5 & 0.5\\
14 & 30 & 30D\('\) & 0/16 & 0/16 & 200000 & 200000 & 1 & --\\
14 & 31 & 31E & 1/16 & 0/16 & 193633 & 200000 & 1 & 1\\
14 & 32 & 32E & 1/16 & 0/16 & 193365 & 200000 & 1 & 1\\
14 & 33 & 33E & 1/16 & 0/16 & 193143 & 200000 & 1 & 1\\
14 & 34 & 34E & 0/16 & 0/16 & 200000 & 200000 & 1 & --\\
14 & 35 & 35E & 0/16 & 0/16 & 200000 & 200000 & 1 & --\\
14 & 36 & 36F & 0/16 & 0/16 & 200000 & 200000 & 1 & --\\
14 & 37 & 37D & 0/16 & 0/16 & 200000 & 200000 & 1 & --\\
14 & 38 & 38D & 3/16 & 0/16 & 172114 & 200000 & 0.25 & 0.25\\
14 & 39 & 39C & 4/16 & 0/16 & 155633 & 200000 & 0.12 & 0.12\\
14 & 40 & 40C & 6/16 & 0/16 & 136340 & 200000 & 0.031 & 0.031\\
14 & 41 & 41E & 0/16 & 0/16 & 200000 & 200000 & 1 & --\\
14 & 42 & 42D & 0/16 & 0/16 & 200000 & 200000 & 1 & --\\
14 & 43 & 43D & 0/16 & 0/16 & 200000 & 200000 & 1 & --\\
14 & 44 & 44D & 0/16 & 0/16 & 200000 & 200000 & 1 & --\\
14 & 45 & 45E & 0/16 & 0/16 & 200000 & 200000 & 1 & --\\
14 & 46 & 46D & 1/16 & 0/16 & 188256 & 200000 & 1 & 1\\
14 & 47 & 47C & 0/16 & 0/16 & 200000 & 200000 & 1 & --\\
14 & 48 & 48C & 0/16 & 0/16 & 200000 & 200000 & 1 & --\\
14 & 49 & 49C & 0/16 & 0/16 & 200000 & 200000 & 1 & --\\
14 & 50 & 50D & 0/16 & 0/16 & 200000 & 200000 & 1 & --\\
14 & 51 & 51E & 0/16 & 0/16 & 200000 & 200000 & 1 & --\\
14 & 52 & 52E & 0/16 & 0/16 & 200000 & 200000 & 1 & --\\
14 & 53 & 53C & 0/16 & 0/16 & 200000 & 200000 & 1 & --\\
14 & 54 & 54C & 0/16 & 0/16 & 200000 & 200000 & 1 & --\\
14 & 55 & 55C & 0/16 & 0/16 & 200000 & 200000 & 1 & --\\
14 & 56 & 56D & 0/16 & 0/16 & 200000 & 200000 & 1 & --\\
14 & 57 & 57C & 0/16 & 0/16 & 200000 & 200000 & 1 & --\\
14 & 58 & 58C & 1/16 & 0/16 & 187848 & 200000 & 1 & 1\\
14 & 59 & 59E & 0/16 & 0/16 & 200000 & 200000 & 1 & --\\
14 & 60 & 60F & 0/16 & 0/16 & 200000 & 200000 & 1 & --\\
14 & 61 & 61C & 0/16 & 0/16 & 200000 & 200000 & 1 & --\\
14 & 62 & 62G & 0/16 & 0/16 & 200000 & 200000 & 1 & --\\
14 & 63 & 63C & 0/16 & 0/16 & 200000 & 200000 & 1 & --\\
14 & 64 & 64C & 0/16 & 0/16 & 200000 & 200000 & 1 & --\\
14 & 65 & 65G & 0/16 & 0/16 & 200000 & 200000 & 1 & --\\
14 & 66 & 66F & 0/16 & 0/16 & 200000 & 200000 & 1 & --\\
14 & 67 & 67E & 0/16 & 0/16 & 200000 & 200000 & 1 & --\\
14 & 68 & 68H & 0/16 & 0/16 & 200000 & 200000 & 1 & --\\
14 & 69 & 69F & 0/16 & 0/16 & 200000 & 200000 & 1 & --\\
14 & 70 & 70F & 0/16 & 0/16 & 200000 & 200000 & 1 & --\\
14 & 71 & 71D & 2/16 & 0/16 & 182526 & 200000 & 0.5 & 0.5\\
14 & 72 & 72E & 1/16 & 0/16 & 188477 & 200000 & 1 & 1\\
14 & 73 & 73E & 1/16 & 0/16 & 188263 & 200000 & 1 & 1\\
14 & 74 & 74C & 3/16 & 0/16 & 167724 & 200000 & 0.25 & 0.25\\
14 & 75 & 75C & 2/16 & 0/16 & 180467 & 200000 & 0.5 & 0.5\\
14 & 76 & 76F & 4/16 & 0/16 & 166634 & 200000 & 0.12 & 0.12\\
14 & 77 & 77D & 1/16 & 0/16 & 190612 & 200000 & 1 & 1\\
14 & 78 & 78F & 1/16 & 0/16 & 192454 & 200000 & 1 & 1\\
14 & 79 & 79F & 0/16 & 0/16 & 200000 & 200000 & 1 & --\\
14 & 80 & 80G & 1/16 & 0/16 & 190366 & 200000 & 1 & 1\\
\end{longtable}
\endgroup

\bibliographystyle{unsrtnat}
\bibliography{refs,refs_jcc,shao2004}
\end{document}